\documentclass[acmsmall]{acmart}
\usepackage{forest}
\usepackage{adjustbox}
\usepackage{rotating}
\usepackage{geometry}
\usepackage{calc}
\usepackage{booktabs}
\usepackage{xurl}
\usepackage{url} 
\usepackage{tcolorbox}
\usepackage{pgfplots}
\pgfplotsset{compat=1.18}
\usepackage{tikz}
\usepackage{tabularx, booktabs}
\usepackage{float}
\usepackage[utf8]{inputenc}
\usepackage[T1]{fontenc}
\usepackage{fontawesome5}
\usetikzlibrary{arrows.meta, positioning, shadows.blur, calc, fit, backgrounds}
\newtcolorbox{commentbox}{
  colback=blue!4,
  colframe=blue!50!black,
  boxrule=0.4pt,
  arc=2pt,
  left=4pt,
  right=4pt,
  top=2pt,
  bottom=2pt,
  width=\linewidth,
  fontupper=\itshape\small
}

\definecolor{primaryColor}{HTML}{005b96} 
\definecolor{accentColor}{HTML}{33a1fd}  
\definecolor{fillColor}{HTML}{ffffff}    

\definecolor{phase1bg}{HTML}{BBDEFB}     
\definecolor{phase2bg}{HTML}{C8E6C9}     
\definecolor{phase3bg}{HTML}{FFE0B2}     

\newcommand{\nodecontent}[3]{%
    \vspace{6pt}
    \textcolor{accentColor}{\huge #1}\\[10pt] 
    \textbf{\large #2}\\[6pt]        
    \footnotesize\textcolor{darkgray}{#3}
    \vspace{2pt}
}

\AtBeginDocument{%
  }

\setcopyright{acmlicensed}
\copyrightyear{2018}
\acmYear{2018}
\acmDOI{XXXXXXX.XXXXXXX}

\acmJournal{JACM}
\acmVolume{37}
\acmNumber{4}
\acmArticle{111}
\acmMonth{8}

\usepackage[dvipsnames]{xcolor}

\begin{document}

\title{Characterizing Faults in Agentic AI: A Taxonomy of Types, Symptoms, and Root Causes}

\author{Mehil B Shah}
\email{shahmehil@dal.ca}
\affiliation{%
  \institution{Dalhousie University}
  \city{Halifax}
  \state{NS}
  \country{Canada}
}

\author{Mohammad Mehdi Morovati}
\email{mehdi.morovati@polymtl.ca}
\affiliation{%
  \institution{Polytechnique Montreal}
  \city{Montreal}
  \state{QC}
  \country{Canada}
}

\author{Mohammad Masudur Rahman}
\email{masud.rahman@dal.ca}
\affiliation{%
  \institution{Dalhousie University}
  \city{Halifax}
  \state{NS}
  \country{Canada}
}

\author{Foutse Khomh}
\email{foutse.khomh@polymtl.ca}
\affiliation{%
  \institution{Polytechnique Montreal}
  \city{Montreal}
  \state{QC}
  \country{Canada}
}

\renewcommand{\shortauthors}{Shah et al.}

\begin{abstract}
Agentic AI systems combine LLM-based reasoning, orchestration, tool invocation, and interaction with external environments. These systems introduce faults that are difficult to characterize using existing taxonomies. To address this gap, we present an empirical study of faults in agentic AI systems. We collected 13,602 issues and pull requests from 40 repositories and, using stratified sampling, selected 385 faults for analysis. Through grounded theory, we derived taxonomies of fault types, symptoms, and root causes. We then used Apriori-based association rule mining to identify relationships among faults, symptoms, and root causes, and validated the taxonomy through a developer study with 145 practitioners. Our analysis produced a taxonomy of 34 fault types, organized into four architectural dimensions. These faults manifested as failures in structured-output interpretation, tool calls, runtime execution, and exception handling, with root causes including data schema mismatches, dependency drift, state management complexity, and model interface instability. Furthermore, association rules showed recurring cross-component propagation, linking structured data, dependency, and state management faults to their symptoms and root causes. Practitioners considered the taxonomy representative of agentic AI failures and suggested refinements related to multi-agent coordination and observability. These findings provide an empirical basis for diagnosing faults and improving reliability in agentic AI systems.
\end{abstract}



\received{20 February 2007}
\received[revised]{12 March 2009}
\received[accepted]{5 June 2009}

\maketitle
\section{Introduction}
Recent years have seen growing interest in building more capable AI systems that can do more than simply engaging in passive conversation. This shift has led to the emergence of \emph{agentic AI systems} that can plan actions and use external tools to achieve complex goals~\cite{xi2025rise, wang2024survey}. Industry increasingly deploys these agentic AI systems in important domains such as autonomous software engineering, enterprise automation, robotics, and real-world decision support~\cite{rahardja2025can, lu2025, liu2023agentbench}. Unlike conversational LLM systems, agentic systems execute multi-step reasoning, maintain internal state across interactions, and actively interact with external environments to accomplish their tasks. This operational independence allows us to automate routine tasks more effectively, improve efficiency in complex workflows, and achieve more scalable system behaviour; at the same time, it introduces substantial challenges for reliability, correctness, and safety.

Because agentic systems differ fundamentally from both traditional deterministic software and LLM-based conversational systems, they may manifest their failures differently. In traditional software and conversational LLM systems, failures typically arise from faulty code or model hallucinations, respectively~\cite{zhang2025llm, lin2025llm, hamill2009common}. In contrast, in agentic systems, failures arise not only from faulty code or model hallucinations, but also from agent orchestration, evolving internal state, and interactions with environmental feedback~\cite{cemri2025multi}. An agent may, for example, generate a seemingly reasonable sequence of actions, yet misinterpret tool outputs, become trapped in infinite reasoning cycles, or maintain incorrect or outdated assumptions about the external file-system state~\cite{he2025plan}. Recent industry reports and security advisories document incidents in which agentic systems deleted critical data, failed to terminate processes, or hallucinated API parameters, resulting in security vulnerabilities and economic losses~\cite{owaspLLM2025}. These failures exacerbate the existing challenges and costs of poor quality software and bugs (e.g., the US economy loses \$2.4T per year)~\cite{crosby2022softwarequality,nist2002softwarebugs}. Beyond these economic consequences, agentic systems introduce risks that extend to physical safety and system integrity. When deployed in safety-critical domains such as healthcare, transportation, industry, and energy infrastructure, failures involving misinterpreted tool outputs, uncontrolled action execution, or agents operating on incorrect system state can directly lead to physical harm, cascading system failures, or loss of life~\cite{aisafetyreport}. Without a clear understanding of how these failures arise, manifest, and propagate through agentic workflows, deploying such systems in real-world environments is inherently risky and, in extreme cases, potentially fatal~\cite{aisafetyreport, amodei2016concrete}. These concerns have prompted growing attention from both industry and academia toward the systematic study of agentic system failures.

Several prior studies have investigated failures in agentic systems. Cemri et al.~\cite{cemri2025multi} analyse failures in multi-agent LLM systems and emphasize coordination challenges, but their analysis overlooks the mapping between agentic failures and system components. Rahardja et al.~\cite{rahardja2025can} study software engineering agents, evaluate their capability to generate patches against issue reports, and propose a taxonomy derived from GitHub issues. However, their taxonomy is focused on guiding automated repair workflows and benchmarking patch-generation success, and thus fails to explain how faults arise within specific agent components or how they propagate through different components. Lu et al.~\cite{lu2025} study autonomous agents in the software engineering domain, analysing failures in programming tasks such as web crawling and data analysis. Using the failure logs, they propose a taxonomy of agent faults. While these advancements are inspiring, the lack of a connection between failures and components in agentic systems limits the effectiveness of the existing taxonomies. In other words, establishing such links is essential to enable systematic diagnosis of agentic failures and the development of appropriate debugging techniques.

In this paper, we conduct a large-scale empirical study of faults in real-world agentic AI systems and examine \textit{where}, \textit{how}, and \textit{why} they arise. Prior studies have largely analysed faults at the task or behavioural level; in contrast, our work adopts a component-grounded perspective that explicitly links agentic faults to system components and identifies the underlying causes of these faults and their observable symptoms. We design our study in five steps as follows. \emph{First}, we construct a dataset of 13{,}602 documented faults from 40 real-world agentic systems. \emph{Second}, we apply stratified sampling to select 385 faults for in-depth manual analysis, preserving representativeness across repository types (e.g., frameworks, applications). \emph{Third}, we use a grounded theory approach~\cite{glaser2017discovery} to inductively derive taxonomies of fault types, symptom patterns, and root causes through iterative coding and comparison. \emph{Fourth}, we apply Apriori-based association rule mining~\cite{agarwal1994fast} to identify high-confidence associations among fault types, symptoms, and root causes, revealing fault propagation patterns. \emph{Finally}, we validate the resulting taxonomies through a structured developer study with 145 participants experienced in agentic system development, establishing ecological validity and practical utility. We also answer the following research questions (RQs) in our study:

\textbf{RQ1:} \emph{What are the characteristics, observable symptoms, and root causes of faults and failures in real-world agentic AI systems?}

Prior analyses of failures in agentic AI systems have largely focused on task outcomes or high-level behavioural errors. Advancing the understanding of agentic AI failures, therefore, requires a systematic identification of fault types, observable symptoms, and root causes. By analysing 385 representative faults, we identify 34 fault types organized into 14 categories and four architectural dimensions, along with 12 symptom categories and 12 root cause categories. Faults occurring in components responsible for reasoning, orchestration, tool interaction, and execution environments indicate that failures emerge across multiple layers of the agent pipeline. Dominant root causes—\emph{Data Schema Mismatches} (28.0\%) and \emph{Dependency Drift} (21.9\%)—reveal systematic contract violations between probabilistically generated artifacts and the deterministic interface, type, and runtime constraints enforced by external software systems.

\textbf{RQ2:} \emph{What statistically significant associations do exist among faults, symptoms, and root causes in agentic AI systems?}

Existing studies typically define failure categories~\cite{cemri2025multi, rahardja2025can, lu2025} without examining how faults originate in agent components and manifest as symptoms in the system. As a result, the mechanisms by which different faults propagate through agentic AI systems remain poorly understood. Using association rule mining, we identify high-confidence associations among fault types, symptoms, and root causes. For example, authentication-related request failures are strongly associated with fragile token refresh mechanisms that mishandle credential expiration, while incorrect time values typically stem from inappropriate datetime conversions. Taken together, these patterns suggest that many faults do not fail transparently; instead, weak error handling and limited logging obscure their underlying causes, turning simple implementation mistakes into difficult-to-diagnose faults.

\textbf{RQ3:} \emph{To what extent do the derived taxonomies of faults, symptoms, and root causes align with the practical experiences of agentic system developers?}

To assess ecological validity beyond offline analysis, we investigate whether the derived taxonomies reflect how practitioners characterize and reason about failures in real agentic systems. We conducted a structured developer survey in which practitioners rated the practical relevance of each fault type based on their experience and provided open-ended feedback on missing or mischaracterised failures. The developer study suggests strong practical alignment, with a mean relevance rating of 3.97 out of 5, indicating that most fault types were viewed as relevant in practice, a large majority of ratings (74.9\%) at 4 or higher reflecting broad endorsement, and high internal consistency (Cronbach’s $\alpha = 0.91$) indicating coherent and stable evaluations across fault types. Most respondents (83.8\%) reported that the taxonomy covered the faults they had encountered, while qualitative feedback suggested refinements (e.g., semantic failures and multi-agent coordination issues). Together, these findings complement the taxonomy by validating its core coverage while identifying targeted refinements that improve its completeness and practical applicability.

Thus, we make the following contributions in the paper:

(a) An empirically grounded, hierarchical taxonomy of faults, symptoms, and root causes, organised into four high-level dimensions aligned with agentic components.

(b) A fault propagation analysis using Apriori-based association rule mining that reveals high-confidence associations among fault types, symptoms, and root causes.

(c) A developer study with 145 participants, establishing the ecological validity and practical utility of the derived taxonomy.

The remainder of the paper is organised as follows. Section~2 describes the methodology, including dataset construction, stratified sampling, grounded theory analysis, association rule mining, and the developer study design. Section~3 presents our findings by answering the three research questions, including the derived taxonomies and high-confidence associations among faults, symptoms, and root causes. Section~4 discusses the implications for debugging, reliability engineering, and agentic system design. Section~5 outlines threats to validity. Section~6 reviews related work, and Section~7 concludes the paper.

\section{Methodology}
\label{sec:method}

Figure~\ref{fig:methodology} presents a schematic overview of our study workflow. We describe each step below.

\begin{figure}[t]
    \centering
    \begin{tikzpicture}[
    font=\sffamily,
    node distance=1.8cm and 1cm,
    process/.style={
        rectangle,
        draw=primaryColor,
        thick,
        fill=fillColor,
        text width=3.4cm,
        minimum height=3.2cm,
        align=center,
        rounded corners=5pt,
        blur shadow={shadow blur steps=5, shadow xshift=1pt, shadow yshift=-1pt}
    },
    arrow/.style={
        line width=2pt,
        -{Latex[length=3.5mm, width=3mm]},
        color=primaryColor,
        rounded corners=5pt
    },
    stepnum/.style={
        circle,
        draw=primaryColor,
        thick,
        fill=white,
        text=primaryColor,
        font=\bfseries\small,
        inner sep=0pt,
        minimum size=7mm,
        blur shadow={shadow blur steps=2, shadow opacity=40}
    },
    group container/.style={
        draw=none,
        inner sep=12pt,
        rounded corners=8pt
    },
    phase label/.style={
        font=\bfseries\small,
        color=darkgray,
        anchor=north
    }
]

    \node (step1) [process] {
        \nodecontent{\faGithub}{Repository Selection}{Filtering \&\\Manual Annotation}
    };
    \node [stepnum] at (step1.north west) {1};

    \node (step2) [process, right=of step1] {
        \nodecontent{\faFilter}{Issue Filtering}{Heuristic \&\\GPT-4.1 Refining}
    };
    \node [stepnum] at (step2.north west) {2};

    \node (step3) [process, right=of step2] {
        \nodecontent{\faLayerGroup}{Dataset \\ Construction}{Stratified Sampling}
    };
    \node [stepnum] at (step3.north west) {3};

    \node (step4) [process, below=of step3] {
        \nodecontent{\faSitemap}{Taxonomy Construction}{Grounded Theory\\(Open/Axial/Sel.)}
    };
    \node [stepnum] at (step4.north west) {4};

    \node (step5) [process, left=of step4] {
        \nodecontent{\faProjectDiagram}{Apriori Analysis}{Association\\Rule Mining}
    };
    \node [stepnum] at (step5.north west) {5};

    \node (step6) [process, left=of step5] {
        \nodecontent{\faClipboardCheck}{Validation}{Developer\\Survey}
    };
    \node [stepnum] at (step6.north west) {6};

    \draw [arrow] (step1) -- (step2);
    \draw [arrow] (step2) -- (step3);
    \draw [arrow] (step3) -- (step4);
    \draw [arrow] (step4) -- (step5);
    \draw [arrow] (step5) -- (step6);

    \begin{scope}[on background layer]
        \node [group container, fill=phase1bg, fit=(step1) (step3)] (group1) {};
        \node [phase label, anchor=south, color=darkgray] at (group1.north) {PHASE I: DATA COLLECTION};

        \node [group container, fill=phase2bg, fit=(step4) (step5)] (group2) {};
        \node [phase label, yshift=-5pt] at (group2.south) {PHASE II: ANALYSIS};

        \node [group container, fill=phase3bg, fit=(step6)] (group3) {};
        \node [phase label, yshift=-5pt] at (group3.south) {PHASE III: VALIDATION};
    \end{scope}

\end{tikzpicture}
    \caption{Schematic overview of our empirical study workflow.}
    \label{fig:methodology}
    \vspace{-0.5cm}
\end{figure}

\subsection{Repository Selection}
\label{subsec:repo-selection}

To curate a representative, high-quality dataset of agentic AI repositories, we followed the repository selection methodology from prior work~\cite{rahardja2025can}. We queried GitHub for repositories associated with the term ``AI agents'' that were active as of June~2025. To focus on robust, widely used systems, we retained repositories with more than 1{,}000 stars and at least 30 issues. These thresholds target projects with sustained maintenance and community engagement, filtering out transient or inactive repositories. This screening yielded 82 candidates.

We then applied a language constraint to align with the dominant agent-development ecosystem: because Python is the predominant language for agentic workflows~\cite{hasan2025empirical}, we excluded repositories implemented in other languages. This reduced the candidate set from 82 to 47.

After automated filtering, we conducted manual annotation to ensure that the dataset comprised only agentic AI frameworks, libraries, or tools. To assess inter-rater agreement, we used Cohen’s Kappa~\cite{diaz2023applying}, a standard measure in similar studies~\cite{morovati2023bugs}. In the first round, two annotators independently labelled all repositories, yielding a Cohen’s Kappa of 0.32 (fair agreement)~\cite{mchugh2012interrater}. The annotators then reconciled disagreements and refined inclusion criteria. A second round produced a Cohen’s Kappa of 0.83 (substantial agreement). This process resulted in a final dataset of 40 Agentic AI repositories.

\subsection{Filtering the Issues}
\label{subsec:issue-filtering}

After selecting repositories, we curated a clean set of closed issues and their corresponding merged pull requests. Closed issues provide evidence of observed failure symptoms and execution contexts, while merged pull requests supply validated fault-localisation and repair information. This pairing enables analysis of both the manifestation of failure and its underlying root causes.  We first applied keyword-based filtering to all closed issues and merged PRs, following established approaches~\cite{abidi2021multi, morovati2024bug}, yielding 24{,}193 candidate items. Following repository mining practices~\cite{kalliamvakou2014promises}, we excluded items written in languages other than English (a small fraction of items were in Chinese~\cite{munaiah2017curating}). We also removed items associated with example or test sections of repositories and entries that did not involve Python files or documentation~\cite{kalliamvakou2014promises}. Applying these criteria reduced the dataset to 19{,}947 candidate closed issues and merged PRs. 

To assess the reliability of the automated filtering, two of the authors manually annotated a representative sample of items (i.e., 68 issues/PRs) from the dataset above. This assessment revealed that 68.6\% of the sampled entries were truly related to a fault, while the remaining 32.4\% constituted noise or non-bug-related items. They included user mistakes (e.g., incorrect environment setup) (e.g., \textit{crewAIInc/crewAI\#1867}), non-reproducible issues (e.g., \textit{microsoft/autogen\#4063}), and feature requests (e.g., \textit{mindsdb/mindsdb\#4721}). Although the initial filtering improved dataset quality, the remaining noise indicated the need for an additional refinement step.

To reduce noise in our dataset, we used GPT-4.1~\cite{gpt4.1} as a final filtering step, consistent with prior work~\cite{li2025rise}. First, we validated GPT-4.1 against a ground-truth set of 68 manually labelled items; it achieved 83\% accuracy and 97\% recall in detecting bug-related issues and PRs. We then manually reviewed a separate sample of 68 items and found GPT-4.1’s judgments matching human interpretation in 85.29\% of cases. We noted that non-bug items included enhancements (e.g., \textit{deepset-ai/haystack\#1453}), stale reports (e.g., \textit{eosphoros-ai/DB-GPT\#1274}), and irreproducible issues (e.g., \textit{Arize-ai/phoenix\#7590}). Using GPT-4.1, we thus removed all items classified as non-bug-related, yielding 13{,}602 closed issues and merged PRs for our analysis. This multi-stage process combined human annotation with AI-assisted validation to produce a high-quality dataset of bug-related activities in agentic AI repositories.

\subsection{Dataset Construction}
\label{subsec:dataset-construction}
To reflect the diversity of the agentic AI ecosystem and support generalisability, we constructed a corpus of 40 repositories through a category-agnostic search process, screening each candidate for relevance to agentic systems without imposing architectural constraints during discovery. Once the final set of repositories was established, we conducted open coding to classify each repository according to its primary architectural role. This classification was guided by the IEEE Standard Glossary of Software Engineering Terminology and established repository taxonomies~\cite{ieeestandards, kim2021denchmark}. This process yielded four categories---\textit{Frameworks}, \textit{Applications}, \textit{Libraries}, and \textit{Tools}---which were used solely to characterise ecosystem coverage rather than to guide repository selection or issue retrieval. The Cohen's kappa for the process was 0.86, which indicates near-perfect agreement~\cite{mchugh2012interrater}.

Because our filtered corpus comprises 13,602 items, conducting a full manual analysis was infeasible. We therefore employed stratified random sampling to construct a statistically representative subset. The target sample size was set to 385, which yields a 95\% confidence level with a 5\% margin of error, a commonly accepted threshold in empirical software engineering studies~\cite{shah2025towards}.

To preserve the underlying distribution, we applied stratified sampling across the four categories so that each category’s share in the sample matched its share in the population. The final dataset contains 385 faults (issues + PRs): \textbf{234} from Frameworks, \textbf{75} from Libraries, \textbf{39} from Tools, and \textbf{37} from Applications.

\subsection{Grounded Theory and Taxonomy Development}
\label{subsec:grounded-theory}

To develop a comprehensive taxonomy of faults in agentic AI systems, we employed a grounded theory approach~\cite{glaser2017discovery} for our analysis. This inductive approach derives categories directly from empirical data rather than relying on an existing taxonomy. We manually examined all 385 sampled items in detail, focusing on reported fault types, observable symptoms, and documented root causes.

To reduce subjectivity and maintain consistency, the analysis was conducted iteratively by the first and second authors. We reviewed the items in batches of 25 and met after each batch to clarify ambiguous cases, resolve disagreements, and refine shared coding guidelines. This process allowed the coding scheme to evolve as new agent-specific fault modes emerged, such as errors from autonomous reasoning loops (e.g., \textit{microsoft/autogen\#4020}) or failures in LLM integration pathways (e.g., \textit{microsoft/autogen\#4770}).

We developed the taxonomy through a three-stage coding process as follows:
\begin{itemize}
    \item \textbf{Open Coding:} We assigned descriptive labels to the raw artefacts, including issue descriptions, execution logs, stack traces, and corresponding fix commits. These labels captured fine-grained technical phenomena, such as \textit{API parameter mismatches} (e.g., \textit{letta-ai/letta\#517}), \textit{token-limit overflows} (e.g., \textit{microsoft/autogen\#4770}), and \textit{dependency version conflicts} (e.g., \textit{crewAIInc/crewAI\#1877}).
    \item \textbf{Axial Coding:} We then examined relationships among these labels to identify conceptual connections and shared underlying causes. Related open codes were grouped into broader sub-categories that reflected the operation of agentic AI systems. For example, LLM configuration and API token-handling faults were consolidated under LLM integration problems, while data-type and encoding-related faults were grouped as interpretive errors.
    \item \textbf{Selective Coding:} Finally, we synthesised these sub-categories into higher-level dimensions aligned with major architectural components of agentic AI systems, such as the cognitive control loop, the tool and environment interaction layer and the runtime environment.
\end{itemize}

Through this process, we distilled 385 individual faults into a coherent taxonomy. The resulting structure captures both low-level technical faults and broader architectural issues, reflecting distinctive failure modes ranging from breakdowns in internal reasoning to integration errors and vulnerabilities in the operating environment.

\subsection{Apriori Analysis}
\label{subsec:apriori}

To identify systematic relationships among fault types, observable symptoms, and root causes, we used the Apriori algorithm~\cite{agarwal1994fast}. Apriori is a widely used data mining technique that finds frequent itemsets in transactional data and generates association rules describing how attributes co-occur. In a rule, the \textit{antecedent} denotes the observed condition(s), and the \textit{consequent} denotes the attribute(s) that tend to co-occur with those conditions within the same fault instances. In our context, the antecedent can be a fault type or a symptom, and the consequent can be a symptom or a root cause. For example, a rule of the form 
\textit{Symptom = Agent Behaviour Anomalies} $\rightarrow$ 
\textit{Root Cause = Configuration Ambiguity} 
indicates that faults exhibiting anomalous agent behaviour frequently have this specific root cause in our dataset.

We encoded each of the 385 analysed faults as a transaction with three categorical attributes: \textit{Fault Type}, \textit{Symptom}, and \textit{Root Cause}. For example, the following transaction indicates that an \textit{LLM Configuration} fault can manifest as \textit{LLM Interaction Errors} and is caused by a \textit{Configuration Ambiguity} (e.g., \textit{deepset-ai/haystack\#4944}):

\[
\begin{aligned}
\textit{Fault Type} &= \text{LLM Configuration Error}, \\
\textit{Symptom} &= \text{LLM Interaction Errors}, \\
\textit{Root Cause} &= \text{Configuration Ambiguity}.
\end{aligned}
\]

To quantify the strength and reliability of each rule, we computed standard association metrics~\cite{dasseni2001hiding}:
\begin{itemize}
    \item \textbf{Support} measures how frequently items in the antecedent and the consequent co-occur in the dataset. For a rule $X \rightarrow Y$, support is defined as the proportion of transactions in which both $X$ and $Y$ appear together.
    \item \textbf{Confidence} measures how likely the consequent is to occur when the antecedent is present. Formally, for a rule $X \rightarrow Y$, confidence is the conditional probability that a transaction containing $X$ also contains $Y$.
\end{itemize}

By filtering for high-confidence rules, we extracted a concise set of associations that summarise relationships among faults, symptoms, and root causes. For instance, a high-confidence rule such as \textit{Symptom = Execution Loop} $\rightarrow$ \textit{Root Cause = Incorrect Termination Condition} suggests that, in our dataset, faults that exhibit execution loops are frequently associated with termination misconfigurations. While these rules do not establish causality, they provide empirical evidence to guide root-cause investigation by highlighting high-confidence symptom–cause associations. These rules provide a systematic way to infer likely causes from observed symptoms, improving the interpretability and practical utility of the taxonomy.

\subsection{Developer Study}
\label{subsec:developer-study}

To assess the comprehensiveness of our proposed taxonomy, we conducted a structured survey with practitioners actively developing agentic AI systems. The survey evaluates whether the taxonomy captures the real-world faults that they routinely encounter.

\subsubsection{Instrument Design}
\label{subsubsec:instrument}

To gather structured feedback from the participants, we designed an online questionnaire targeting different aspects of our proposed taxonomy. The survey was divided into three sections. The first section collected \textit{demographic information}, including participants’ current professional role, total years of experience in software engineering, and experience with agentic AI frameworks such as AutoGen~\cite{autogen}, LangChain~\cite{langchain}, and BabyAGI~\cite{babyagi}. This information allowed us to assess respondents’ expertise and ensure that feedback came from individuals with relevant experience.

The second section focused on \textit{taxonomy evaluation}. Participants reviewed the full hierarchical taxonomy and rated the practical relevance of the 34 third-level fault types on a five-point Likert scale. We then aggregated ratings to examine results at higher levels of the taxonomy hierarchy. This provided fine-grained quantitative evidence of alignment between the taxonomy and practitioners' experience of agentic faults. The final section collected \textit{open-ended feedback} on completeness, inviting participants to describe missing faults, symptoms, or root causes and suggest additions or refinements.

Before administering the main survey, we conducted a pilot study with two academic researchers and two industry practitioners. Feedback from the pilot led to refinements in category definitions to reduce ambiguity and improve interpretability. For example, we revised several category descriptions to include clearer boundary conditions and short illustrative examples, ensuring that similar but distinct concepts were more explicitly differentiated. We estimated the survey would take approximately 10--15 minutes.

\subsubsection{Participant Selection}
\label{subsubsec:participants}

To ensure high-quality and relevant responses, we employed targeted recruitment. First, we identified active contributors to the selected repositories. Using public commit histories, we extracted the public email addresses of top contributors and sent personalised invitations. This ensured participants were familiar with practical challenges in agentic AI development.

Second, to broaden reach, we disseminated the survey through online communities frequented by AI developers, including specialised Discord servers and Reddit forums dedicated to LLMs and autonomous agents. This combination of direct recruitment and community outreach helped us capture feedback from individuals with relevant experience.

\subsubsection{Demographics}
\label{subsubsec:demographics}

A total of 145 participants completed the survey, representing diverse professional backgrounds, experience levels, and expertise in frameworks. By affiliation, 76.1\% were from industry and 23.9\% from academia. Among them, 29.0\% reported more than 5 years of experience in software engineering, and 47.8\% reported at least 1 year of experience in agentic AI systems.

Participants reported familiarity with a range of agentic AI frameworks. The most common were LangChain (69.6\%) and AutoGen (34.8\%), with additional experience across other frameworks. This distribution indicates broad cross-framework familiarity, providing a solid basis for evaluating our taxonomy’s completeness, accuracy, and usability.

\subsection{Analysis of Developer Study Results}
\label{subsec:survey-analysis}

We analysed the developer study responses using complementary quantitative and qualitative methods to evaluate the taxonomy systematically.

\paragraph{Quantitative analysis.}
Participants rated the practical relevance of each fault type on a five-point Likert scale (1 = not relevant, 5 = highly relevant). For each fault, we computed descriptive statistics (mean, median, and standard deviation) and agreement proportions (ratings $\geq 4$). We aggregated ratings at subcategory, category, and dimension levels to examine consistency across the hierarchical structure. To assess deviation from neutrality, we compared mean ratings against the midpoint of the scale (3.0) using one-sample tests on aggregated ratings. We also examined rating distributions across experience strata (e.g., years of agentic AI development) to assess stability across respondent groups. We evaluated internal consistency of ratings across taxonomy elements using Cronbach’s $\alpha$.

\paragraph{Qualitative analysis.}
We analysed the open-ended responses to identify limitations in taxonomy coverage and specification. Two authors independently reviewed all responses and applied inductive thematic coding to segments describing missing, unclear, or misclassified faults. They compared initial codes, resolved discrepancies through discussion, and iteratively refined a shared coding scheme. We then grouped the agreed codes into higher-level themes representing recurring coverage gaps, definition ambiguities, and structural placement issues within the taxonomy hierarchy. The resulting themes and representative examples are reported in RQ3. The authors resolved remaining disagreements through consensus.

\section{Study Results}
This section presents the findings of our study by answering three research questions.

\begin{figure}[htbp]
\centering
\begin{adjustbox}{max width=\textwidth, max height=0.9\textheight, keepaspectratio}
\begin{forest}
  for tree={
    grow'=0,
    draw,
    rounded corners,
    font=\scriptsize\sffamily,
    align=center,
    inner ysep=2pt,
    edge={-latex},
    l sep=8mm,
    s sep=2pt,
    anchor=west,
    child anchor=west,
    parent anchor=east,
    calign=center
  }
  [{\parbox{3.2cm}{\centering Fault Taxonomy\\for Agentic AI Systems}}, fill=gray!10
    [{\parbox{2.3cm}{\centering I. Agent Reasoning\\\& Control\\(86)}}, fill=cyan!25
      [{\parbox{2.2cm}{\centering 1. LLM\\Interaction\\(48)}}, fill=cyan!15
        [{LLM Configuration Error (13)}, fill=cyan!8]
        [{LLM Invocation Error (28)}, fill=cyan!8]
        [{Token Budget Error (6)}, fill=cyan!8]
        [{LLM Authentication Failure (1)}, fill=cyan!8]
      ]
      [{\parbox{2.2cm}{\centering 2. Agent Lifecycle\\and State\\(38)}}, fill=cyan!15
        [{Agent Step Execution Error (26)}, fill=cyan!8]
        [{Agent State Transition Error (9)}, fill=cyan!8]
        [{Agent Termination Error (3)}, fill=cyan!8]
      ]
    ]
    [{\parbox{2.3cm}{\centering II. Context\\\& Memory\\(67)}}, fill=lime!30
      [{\parbox{2.1cm}{\centering 1. Memory\\Management\\(8)}}, fill=lime!18
        [{Memory Persistence Failure (8)}, fill=lime!10]
      ]
      [{\parbox{2.1cm}{\centering 2. Input\\Interpretation\\(59)}}, fill=lime!18
        [{Structured Data Error (46)}, fill=lime!10]
        [{Constraint Violation (7)}, fill=lime!10]
        [{Representation Encoding Error (4)}, fill=lime!10]
        [{Input Validation Error (2)}, fill=lime!10]
      ]
    ]
    [{\parbox{2.3cm}{\centering III. Tooling, Integration \\\& Actuation (62)}}, fill=pink!35
      [{\parbox{2.1cm}{\centering 1. Tool\\Invocation\\(11)}}, fill=pink!20
        [{Tool Invocation Error (9)}, fill=pink!10]
        [{Tool Configuration Error (2)}, fill=pink!10]
      ]
      [{\parbox{2.1cm}{\centering 2. External\\Access\\(19)}}, fill=pink!20
        [{External Connection Error (6)}, fill=pink!10]
        [{External Authentication Failure (8)}, fill=pink!10]
        [{External Authorization Error (5)}, fill=pink!10]
      ]
      [{\parbox{2.1cm}{\centering 3. Resource\\Interaction\\(21)}}, fill=pink!20
        [{Resource Access Error (16)}, fill=pink!10]
        [{Storage Setup Error (5)}, fill=pink!10]
      ]
      [{\parbox{2.1cm}{\centering 4. System\\Coordination\\(6)}}, fill=pink!20
        [{Synchronization Error (6)}, fill=pink!10]
      ]
      [{\parbox{2.1cm}{\centering 5. Execution\\Monitoring\\(5)}}, fill=pink!20
        [{Execution Monitoring Failure (5)}, fill=pink!10]
      ]
    ]
    [{\parbox{2.3cm}{\centering IV. System Infrastructure\\\& Reliability\\(155)}}, fill=violet!30
      [{\parbox{2.1cm}{\centering 1. Dependency \&\\Environment\\Management\\(72)}}, fill=violet!18
        [{Dependency Compatibility Error (34)}, fill=violet!8]
        [{Module Import Failure (15)}, fill=violet!8]
        [{Dependency Installation Failure (14)}, fill=violet!8]
        [{Package Resolver Error (7)}, fill=violet!8]
        [{Environment Configuration Error (2)}, fill=violet!8]
      ]
      [{\parbox{2.1cm}{\centering 2. Platform \&\\Integration\\Compatibility\\(15)}}, fill=violet!18
        [{Platform Capability Mismatch (8)}, fill=violet!8]
        [{Unsupported Platform (3)}, fill=violet!8]
        [{Component Integration Error (3)}, fill=violet!8]
        [{Missing Function/Interface (1)}, fill=violet!8]
      ]
      [{\parbox{2.1cm}{\centering 3. Failure Handling\\\& Implementation\\Robustness\\(47)}}, fill=violet!18
        [{Exception Handling Failure (27)}, fill=violet!8]
        [{Implementation Logic Error (20)}, fill=violet!8]
      ]
      [{\parbox{2.1cm}{\centering 4. User\\Interface\\Defects\\(13)}}, fill=violet!18
        [{UI Rendering Defect (13)}, fill=violet!8]
      ]
      [{\parbox{2.1cm}{\centering 5. Documentation\\Issues\\(8)}}, fill=violet!18
        [{Documentation Defect (8)}, fill=violet!8]
      ]
    ]
  ]
\end{forest}
\end{adjustbox}
\caption{Taxonomy of faults in agentic AI systems. Numbers in parentheses indicate the frequency of faults.}
\label{fig:taxonomy}
\end{figure}

\subsection{RQ1: What are the characteristics, observable symptoms, and root causes of bugs and failures in real-world agentic AI systems?}
\label{sec:rq1}

Our analysis of 385 real-world faults demonstrates that failures in agentic AI systems differ fundamentally from those in traditional software systems. Their differences are driven by the architectural demands of autonomy: agents must (i) configure and integrate an LLM-based cognitive core, (ii) maintain long-lived state across iterative control loops, and (iii) invoke external tools and services in dynamic environments. To characterise such a fault landscape, we developed a taxonomy comprising 34 fault types grouped into 14 major categories. We also organised them into four high-level dimensions that correspond to core agent capabilities required for autonomy~\cite{wang2024survey, xi2025rise}.

\subsubsection{Taxonomy of Faults in Agentic AI Systems}
\label{subsec:rq1_fault_taxonomy}

\subsubsection*{I. Agent Reasoning \& Control (86 faults)}
\label{subsec:dim1}

This dimension captures faults intrinsic to an agent’s cognitive architecture and control flows. In contrast to conventional software systems, agentic applications rely on LLM-driven reasoning and orchestration logic that coordinate decision-making, iterative execution, and state updates across multiple steps. Faults in this dimension therefore emerge from the interactions between model behaviour and the control mechanisms governing the agent execution.

\textbf{LLM Interaction Faults (48 faults)}: These faults can be found at the interface between an agent framework and an LLM provider. 
This category includes incorrect model configuration (e.g., selecting an unsupported or incorrect model), incorrect LLM provider configuration (e.g.,  misconfigured endpoints for accessing the model), incompatible API usage, inconsistent token accounting, and errors in context-window handling. Such issues typically occur when frameworks have an incorrect assumption of model capabilities, rely on outdated API schemas, or maintain a wrong tokenizer for their operation. These faults can manifest as rejected API calls, context-window overflows, truncated prompts, incorrect reporting of token usage, or unintended tool invocation during the interaction between the LLM and the agent.

For example, \textit{letta-ai/letta} (e.g., \href{https://github.com/letta-ai/letta/issues/180}{Issue \#180}) specified incorrect file paths in its LLM configuration, resulting in a \textit{FileNotFoundError} during persona template retrieval. API evolution can also introduce incompatibilities: \textit{microsoft/autogen} (e.g., \href{https://github.com/microsoft/autogen/pull/2718}{PR \#2718}) experienced failing assistant tests following an OpenAI API change, while \textit{agno-agi/agno} (e.g., \href{https://github.com/agno-agi/agno/issues/2336}{Issue \#2336}) triggered redundant tool execution due to a mismatch between the agentic framework’s streaming logic and the model provider's streaming API behaviour. Token accounting issues also arise in practice when the input and output tokens in an agent are handled incorrectly; for instance, \textit{microsoft/autogen} (e.g., \href{https://github.com/microsoft/autogen/issues/2702}{Issue \#2702}) reported an incorrect token usage for \textit{gpt-4o} due to mismatches between internal tracking and provider tokenisation. Similarly, \textit{Arize-ai/phoenix} (e.g., \href{https://github.com/Arize-ai/phoenix/pull/5976}{PR \#5976}) provided a wrong token count due to an incorrect counter implementation.

\textbf{Agent Lifecycle and State (38 faults)}: These faults indicate issues in the execution lifecycle of autonomous agents. Agentic systems operate through iterative control loops that schedule actions, delegate tasks, and update internal execution state across multiple steps. Faults in this category include incorrect execution scheduling, improper task delegation, inconsistent state transitions across steps, and defective termination logic. Such defects disrupt the intended orchestration and may cause the agents to skip tasks, repeat actions unexpectedly, lose track of intermediate execution state, or enter uncontrolled execution loops.

For example, \textit{crewAIInc/crewAI} (e.g., \href{https://github.com/crewAIInc/crewAI/issues/1501}{Issue \#1501}) prevented coworker agents from executing due to defects in task delegation logic. In \textit{langflow-ai/langflow} (e.g., \href{https://github.com/langflow-ai/langflow/pull/2594}{PR \#2594}), an incorrect run configuration caused the execution pipeline to run multiple times, leading to redundant builds and duplicate tool calls. Similarly, \textit{microsoft/autogen} (e.g., \href{https://github.com/microsoft/autogen/pull/5352}{PR \#5352}) exhibited inconsistent assistant behaviour due to improper state transitions across turns. Finally, \textit{microsoft/autogen} (e.g., \href{https://github.com/microsoft/autogen/issues/4020}{Issue \#4020}) failed to stop agent execution due to an improper termination logic.

\subsubsection*{II. Context \& Memory (67 faults)}
\label{subsec:dim3}

This dimension represents issues in how agents ingest information, maintain contextual artifacts, and interpret heterogeneous inputs. Agent pipelines frequently combine unstructured model outputs (e.g., regular text) with structured artifacts such as source code, files, or database records. Defects in representation, interpretation, or context persistence can therefore propagate through the agentic pipeline, disrupting downstream reasoning and producing inconsistent behaviour.

\textbf{Memory Management (8 faults)}: These faults arise when mechanisms responsible for storing or retrieving contextual artifacts are incorrectly implemented. They include faults in serialization or deserialization logic, incorrect timestamp management, and poor management of persistent storage. Such defects prevent agents from reliably preserving contextual information across executions or sessions. In practice, these faults lead to corrupted context artifacts, temporal inconsistencies in stored records, or system failures when retrieving previously saved data. 

For example, \textit{camel-ai/camel} (e.g., \href{https://github.com/camel-ai/camel/pull/2119}{PR \#2119}) mishandled timestamp updates during memory writes, introducing temporal inconsistencies in stored interaction history. Similarly, \textit{lavague-ai/LaVague} (e.g., \href{https://github.com/lavague-ai/LaVague/issues/338}{Issue \#338}) generated screenshot filenames whose paths exceeded operating-system limits, preventing state artifacts from being written to disk.

\textbf{Input Interpretation (59 faults)}: These faults arise when inputs to the agent are incorrectly parsed, validated, transformed, or interpreted. Because agent pipelines process heterogeneous data—including model outputs, structured artifacts, and external files—errors in interpretation logic can propagate through the pipeline and compromise downstream processing. Such faults include incorrect assumptions about input structures or types, flawed parsing logic, inconsistent handling of character encodings, and missing validation checks on the provided data. They may also occur when systems incorrectly infer file types or apply incompatible parsing logic to input artifacts. These issues can lead to runtime exceptions, corrupted intermediate representations, malformed outputs, or failed downstream tasks operating on invalid data~\cite{shah2025data}. 

For example, \textit{deepset-ai/haystack} (e.g., \href{https://github.com/deepset-ai/haystack/issues/6098}{Issue \#6098}) crashed when a method returned \textit{None} instead of the expected tuple, causing a runtime failure. Similarly, \textit{deepset-ai/haystack} (e.g., \href{https://github.com/deepset-ai/haystack/pull/2262}{PR \#2262}) failed to extract tabular data from model outputs due to parsing defects in \textit{ParsrConverter}. Encoding inconsistencies can also introduce system failures; for instance, \textit{langflow-ai/langflow} (e.g., \href{https://github.com/langflow-ai/langflow/issues/8482}{Issue \#8482}) failed to download projects containing Cyrillic filenames due to incorrect encoding assumptions. Missing validation checks can also propagate malformed data through the system; for example, \textit{camel-ai/camel} (e.g., \href{https://github.com/camel-ai/camel/issues/1361}{Issue \#1361}) generated invalid arXiv URLs because the returned identifiers were not validated before use.

\subsubsection*{III. Tooling, Integration \& Actuation (62 faults)}
\label{subsec:dim2}

This dimension captures the faults that arise when agents invoke tools, APIs, and external systems. While an agent’s reasoning process is probabilistic, its interactions with external services rely on deterministic interfaces with strict operational constraints. Consequently, defects frequently occur at these system boundaries when the agents' requests, configurations, or execution assumptions do not align with the requirements of external tools and services.

\textbf{Tool Invocation (11 faults)}: These faults arise when agents invoke a defective tool or invoke a tool incorrectly. They include violations of API contracts, unsupported operations, and mismatches between expected and supplied parameters. Such issues prevent agents from interacting correctly with external services even when the intended action might be valid. In practice, they manifest as rejected API calls, malformed request errors, or unsuccessful tool execution. For example, \textit{deepset-ai/haystack} (e.g., \href{https://github.com/deepset-ai/haystack/pull/4703}{PR \#4703}) violated query-builder constraints of \textit{Weaviate BM25} by passing unsupported arguments, resulting in a program crash. Similarly, \textit{SWE-agent/SWE-agent} (e.g., \href{https://github.com/SWE-agent/SWE-agent/issues/1159}{Issue \#1159}) experienced tool-calling failures due to incorrectly structured arguments supplied during tool invocation.

\textbf{External Access (19 faults)}: Faults belonging to this category occur when agents fail to establish or maintain access to external services. They include invalid endpoints, misconfigured connection settings, and poor management of authentication credentials or permissions. Such issues can prevent agents from communicating with external systems or executing authorised operations. For instance, \textit{mindsdb/mindsdb} (e.g., \href{https://github.com/mindsdb/mindsdb/pull/7410}{PR \#7410}) experienced broken Cassandra connections due to an incorrect \texttt{secure\_connect\_bundle} path in its configuration. As another example, the \textit{Arize-ai/phoenix} playground (e.g., \href{https://github.com/Arize-ai/phoenix/pull/5003}{PR \#5003}) crashed due to missing credentials in its request flow, while \textit{browser-use/browser-use} (e.g., \href{https://github.com/browser-use/browser-use/issues/2571}{Issue \#2571}) encountered permission-denied errors when attempting to access hardware peripherals without the required permissions.

\textbf{Resource Interaction (21 faults)}: These faults stem from shared resources required for tool-mediated execution in the agentic AI systems. They include invalid resource paths, incorrect filesystem assumptions, and misconfigured storage or databases. Such issues can prevent tools from accessing required resources or cause failures when reading, writing, or managing stored artifacts. For example, \textit{agno-agi/agno} (e.g., \href{https://github.com/agno-agi/agno/issues/1189}{Issue \#1189}) failed during hybrid search with LanceDB due to incorrect database configuration.

\textbf{System Coordination (6 faults)}: These faults arise from poor coordination of concurrent or asynchronous execution within agent systems. Agents frequently rely on parallel operations, streaming responses, or asynchronous tool calls, which require careful ordering and synchronisation of observable effects. When coordination mechanisms are incorrect or incomplete, race conditions can occur, and output could be incoherent. As an example, \textit{FoundationAgents/MetaGPT} (e.g., \href{https://github.com/FoundationAgents/MetaGPT/pull/1381}{PR \#1381}) produced incoherent streamed outputs due to a poor coordination of parallel response streams. Similarly, \textit{qodo-ai/pr-agent} (e.g., \href{https://github.com/qodo-ai/pr-agent/issues/1324}{Issue \#1324}) experienced CI/CD failures due to incorrect synchronisation during concurrent execution.

\textbf{Execution Monitoring (5 faults)}: These faults emerge when mechanisms intended to record or expose system behaviour are missing or incorrectly implemented. They involve defects in logging, tracing, or metric instrumentation that reduce visibility into system execution and make failures difficult to diagnose. When observability signals are absent or incorrect, developers may be unable to reconstruct the sequence of events leading to a failure. For example, \textit{langflow-ai/langflow} (e.g., \href{https://github.com/langflow-ai/langflow/pull/8457}{PR \#8457}) logged incorrect database metrics, which made it difficult to observe the database operations performed during agent execution.

\subsubsection*{IV. System Infrastructure \& Reliability (155 faults)}
\label{subsec:dim4}

This dimension captures system-level faults that arise from the infrastructure supporting agentic AI systems, including their runtime environments, dependencies, and operational tooling. Although these faults are not intrinsic to agent reasoning or decision-making, they can significantly affect agentic AI systems in their operations.

\textbf{Dependency and Environment Management (72 faults)}: These faults arise from defects in specifying, resolving, or installing software dependencies required for agent execution. They occur when dependency versions are incompatible, required packages are missing, or environment provisioning produces inconsistent installations across machines. Such issues often emerge when dependency specifications become outdated following upstream library changes, when dependency constraints cannot be satisfied by package managers, or when environment provisioning fails due to packaging constraints or system limitations. They may also occur when package metadata (e.g., supported Python versions or platform tags) is incompatible with the target runtime environment. In practice, these faults lead to failures such as unsuccessful environment setup, dependency resolution conflicts, broken functionality after dependency upgrades, or inconsistent behaviour across installations. 

For example, \textit{deepset-ai/haystack} (e.g., \href{https://github.com/deepset-ai/haystack/issues/6652}{Issue \#6652}) experienced failures after an upstream structural change in the dependency specifications. Similarly, \textit{FoundationAgents/MetaGPT} (e.g., \href{https://github.com/FoundationAgents/MetaGPT/issues/1658}{Issue \#1658}) failed during installation due to excessively long file paths that prevented successful environment provisioning. Dependency resolution conflicts can also prevent environment setup; for example, \textit{letta-ai/letta} (e.g., \href{https://github.com/letta-ai/letta/issues/1068}{Issue \#1068}) failed during installation because a dependency was incompatible with the Python version in use.

\textbf{Platform \& Integration Compatibility (15 faults)}: These faults arise when assumptions about the underlying platform are not satisfied at runtime. They include wrong assumptions related to operating systems, CPU architectures, filesystem constraints, or system utilities expected by the application or container environment. Such issues frequently occur when binaries or containers are built for one architecture but executed on another, when required system commands or services are unavailable on the host platform, or when software components assume platform-specific behaviour that does not hold across environments. In practice, these faults lead to failures such as binaries failing to execute on incompatible architectures, containers failing to start due to unsupported host environments, or runtime errors caused by missing system utilities. 

For example, \textit{microsoft/autogen} (e.g., \href{https://github.com/microsoft/autogen/issues/1382}{Issue \#1382}) encountered file-not-found errors when running Docker containers on Apple Silicon because the container environment assumed an x86-based runtime setup.

\textbf{Failure Handling \& Implementation Robustness (47 faults)}: These faults stem from mechanisms responsible for detecting, propagating, and handling failures during agent execution. Because agentic systems frequently interact with multiple external services and execute long-running tasks, reliable error handling is essential to prevent failures from silently disrupting the system behaviour. Such issues occur when exceptions are suppressed, error signals are incorrectly propagated, or failure recovery mechanisms do not work. In practice, these faults manifest as silent execution failures, incomplete task execution, or systems continuing operation with erroneous internal states. For instance, \textit{mindsdb/mindsdb} (e.g., \href{https://github.com/mindsdb/mindsdb/pull/9659}{PR \#9659}) failed to surface errors caused by the deletion operation during database, suppressing critical state information.

\textbf{User Interface Defects (13 faults)}: These faults arise when user-facing interfaces incorrectly represent system state or execution context. Agent systems often expose dashboards, visualisations, or interactive interfaces that allow developers and users to inspect the execution signals from the system. When interface components render incomplete, outdated, or inconsistent information, they can misrepresent system behaviour and hinder any diagnosis. In practice, such defects appear as missing visual updates, incorrect timestamps or data displays, or broken interaction elements. For example, \textit{Arize-ai/phoenix} (e.g., \href{https://github.com/Arize-ai/phoenix/issues/415}{Issue \#415}) failed to render selected timestamps on the embedding details page, preventing users from accurately inspecting recorded execution data.

\textbf{Documentation Issues (8 faults)}: These faults arise when the system documentation inaccurately describes system behaviour, configuration, or usage requirements. Since agent systems often integrate multiple tools, APIs, and configuration parameters, the incorrect documentation can mislead developers. Such issues include incorrect examples, typographical errors in interface specifications, or incomplete explanations of system behaviour. Although they do not directly affect program execution, they can propagate incorrect implementation or operational practices. As an example, in the \textit{browser-use/browser-use} repository (e.g., \href{https://github.com/browser-use/browser-use/pull/2442}{PR \#2442}), the documentation included an incorrect CLI command for using the framework, resulting in installation failures for developers.
\subsubsection{Symptoms of Faults in Agentic AI Systems}

We identified 12 observable symptoms through which the faults manifest themselves in agentic AI systems. These symptoms are developer-facing signals that appear when faults disrupt different components of the agent execution pipeline. Table~\ref{tab:symptom_distribution} summarises their prevalence across the 385 faults in our dataset.

\begin{table}[htbp]
\centering
\caption{Distribution of Fault Symptoms in Agentic Systems}
\label{tab:symptom_distribution}
\begin{tabular}{lrr}
\toprule
\textbf{Symptom Category} & \textbf{Count} & \textbf{Percentage} \\
\midrule
Data and Validation Errors & 80 & 22.7\% \\
Runtime Execution Errors & 68 & 19.3\% \\
Unhandled Exceptions & 45 & 12.8\% \\
Build Errors & 32 & 9.1\% \\
Dependency Installation Errors & 28 & 8.0\% \\
Agent Behaviour Anomalies & 24 & 6.8\% \\
LLM Interaction Errors & 22 & 6.2\% \\
Network Connectivity Errors & 17 & 4.8\% \\
User Interface Errors & 15 & 4.3\% \\
Tool Invocation Errors & 9 & 2.6\% \\
File and Resource Access Errors & 9 & 2.6\% \\
Concurrency Errors & 3 & 0.9\% \\
\bottomrule
\end{tabular}
\end{table}

\textbf{Data \& Validation Errors (80 occurrences)} surface when intermediate artefacts exchanged between components in the agentic reasoning pipeline cannot be interpreted correctly. These symptoms are typically observed in components such as artifact parsers, schema validators, and data transformation modules. Developers encounter parsing failures, schema validation exceptions, malformed intermediate representations, or incompatible values that downstream components cannot process, interrupting reasoning chains or tool invocations. These symptoms commonly arise from faults in the \textit{Input Interpretation} category of our taxonomy, and may correspond to specific faults such as \textit{Structured Data Error} or \textit{Input Validation Error}. For example, in \textit{deepset-ai/haystack} (e.g., \href{https://github.com/deepset-ai/haystack/pull/6382}{PR \#6382}), the execution pipeline failed with a runtime validation error when the \textit{token\_to\_chars} function returned \textit{None}, causing downstream components expecting a tuple to crash.

\textbf{Runtime Execution Errors (68 occurrences)} cause the agent workflows to terminate unexpectedly or to halt. These symptoms are observed in orchestration components that coordinate reasoning steps, tool execution, and workflow control. Developers typically observe interrupted workflows, failed execution steps, or incomplete reasoning loops during task execution. These symptoms commonly arise from faults in the \textit{Agent Lifecycle and State} and \textit{Failure Handling \& Implementation Robustness} categories. For example, in \textit{microsoft/autogen} (e.g., \href{https://github.com/microsoft/autogen/pull/2718}{PR \#2718}), agent execution stopped midway through a task workflow, leaving the reasoning process incomplete.

\textbf{Unhandled Exceptions (45 occurrences)} occur when runtime execution errors propagate through the agentic system without being captured and handled appropriately. These errors can be observed in execution controllers and service handlers that manage agent workflows and tool interactions. Developers typically see abrupt program termination accompanied by stack traces or runtime error messages. These symptoms are commonly associated with faults in \textit{Failure Handling \& Implementation Robustness}. For example, in \textit{Netflix/metaflow} (e.g., \href{https://github.com/Netflix/metaflow/pull/1444}{PR \#1444}), the system terminated with a stack trace after an exception propagated through the execution pipeline without being caught and handled.

\textbf{Build Errors (32 occurrences)} become visible when the components supporting agent capabilities cannot be compiled or packaged successfully. These symptoms are observed in build pipelines and integration workflows that involve LLM integration, agent orchestration, tool execution, and external service interaction. Developers observe compilation failures, unresolved imports, or CI failures when encountering these faults. These symptoms typically correspond to faults in \textit{Dependency \& Environment Management} or \textit{Platform \& Integration Compatibility}. For example, in \textit{microsoft/autogen} (e.g., \href{https://github.com/microsoft/autogen/issues/5715}{Issue \#5715}), the CI pipeline failed during the build stage due to unresolved module imports.

\textbf{Dependency Installation Errors (28 occurrences)} arise when the required libraries supporting core agent capabilities cannot be installed successfully. These symptoms are observed in the modules responsible for environment provisioning and dependency management. Developers encounter unresolved dependency trees, incompatible package versions, or missing modules when installing libraries required for model interaction, memory systems, or tool execution, which can prevent the agent framework from running. These symptoms are commonly associated with faults in \textit{Dependency \& Environment Management}. For example, in \textit{crewAIInc/crewAI} (e.g., \href{https://github.com/crewAIInc/crewAI/issues/1877}{Issue \#1877}), the installation process aborted due to dependency resolution errors involving libraries required for the agent framework, preventing the system from being installed successfully.

\textbf{Agent Behaviour Anomalies (24 occurrences)} become evident when agents produce actions or outputs that deviate from the planned workflow logic. These symptoms are observed in reasoning and orchestration components responsible for decision-making and multi-step workflows in the agentic systems. Developers observe them as repeated reasoning steps, agent role violations, loss of conversational context, or breakdowns in coordination among collaborating agents. Such behaviours often occur without explicit crashes, making them harder to diagnose~\cite{xing2025looking, raheem2025agentic}. These symptoms are typically associated with faults in \textit{Agent Lifecycle and State} and \textit{LLM Integration}. For example, in \textit{openai/openai-agents-python} (e.g., \href{https://github.com/openai/openai-agents-python/issues/968}{Issue \#968}), the agent repeatedly generated responses that violated the provided role instructions during task execution.

\textbf{LLM Interaction Errors (22 occurrences)} arise when agents fail to obtain responses from their language models during the reasoning stage. They can be observed in components responsible for communicating with external model APIs, such as prompt dispatchers and model interface layers. Developers encounter rejected API calls, unsupported model invocations, authentication failures, or context overflows when the agent attempts to generate outputs. These symptoms correspond to faults in the taxonomy category of \textit{LLM Integration}. For instance, in \textit{browser-use/browser-use} (e.g., \href{https://github.com/browser-use/browser-use/pull/1863}{PR \#1863}), agent requests to the language model API were rejected, preventing the agent from generating responses.

\textbf{Network Connectivity Errors (17 occurrences)} occur when agents cannot communicate with external services required for their task execution. These symptoms are observed in connector modules that interact with external APIs, databases, or remote services. Developers observe connection timeouts, unreachable endpoints, or rejected requests in their agentic systems. These symptoms correspond to faults in \textit{External Access}. For example, in \textit{deepset-ai/haystack} (e.g., \href{https://github.com/deepset-ai/haystack/issues/617}{Issue \#617}), requests to the external search service repeatedly timed out, interrupting the agent’s retrieval workflow.

\textbf{User Interface Errors (15 occurrences)} refer to incorrect or incomplete information visualization about agent execution. They can be observed in interface components that visualise execution logs, metrics, and task traces. Developers observe broken interactions, inconsistent timestamps, or missing execution history while encountering these errors. These issues affect system observability rather than execution itself and are typically associated with faults in \textit{User Interface Defects}. For example, in \textit{Arize-ai/phoenix} (e.g., \href{https://github.com/Arize-ai/phoenix/issues/415}{Issue \#415}), the monitoring dashboard displayed incorrect timestamps for recorded embeddings, leading to misleading execution traces.

\textbf{Tool Invocation Errors (9 occurrences)} occur when agents attempt to invoke external functions or services. These symptoms are observed in modules that prepare requests to invoke external tools. Developers observe runtime errors, interrupted workflow steps, or unavailable capabilities when tool calls fail. These disruptions directly interrupt task execution and can lead to incorrect final outcomes. These symptoms correspond to faults in the taxonomy category of \textit{Tool Invocation}. For instance, in \textit{FoundationAgents/MetaGPT} (e.g., \href{https://github.com/FoundationAgents/MetaGPT/issues/334}{Issue \#334}), a tool invocation failed and resulted in a \textit{NotImplementedError}, interrupting the task workflow of the corresponding agent.

\textbf{File and Resource Access Errors (9 occurrences)} occur when agents fail to read or write artefacts required for their execution. These symptoms are observed in components that interact with local files, configuration artefacts, or persistent storage. Developers encounter file-not-found errors, read failures, or unsuccessful attempts to save outputs. These issues disrupt workflows that rely on configuration files, templates, or stored artefacts. These symptoms correspond to the \textit{Resource Interaction} fault. For example, in \textit{letta-ai/letta} (e.g., \href{https://github.com/letta-ai/letta/issues/180}{Issue \#180}), the agent failed to start because the required persona template file could not be located in the expected directory.

\textbf{Concurrency Errors (3 occurrences)} emerge in multi-agent or asynchronous execution settings where multiple components operate simultaneously. These symptoms are observed in coordination mechanisms that synchronise concurrent agent actions and manage asynchronous workflows. Developers observe inconsistent output ordering, merged responses, or non-deterministic behaviour across repeated runs. These symptoms correspond to faults in the taxonomy category of \textit{System Coordination}. For example, in \textit{FoundationAgents/MetaGPT} (e.g., \href{https://github.com/FoundationAgents/MetaGPT/pull/1381}{PR \#1381}), outputs produced by parallel agents were merged incorrectly, resulting in misaligned responses in the final output.
\subsubsection{Root Cause Analysis}
\label{sec:root_cause_analysis}

We also identified 12 root causes that explain why faults occur in agentic AI systems. These root causes give rise to observable symptoms and concrete fault types across different stages of the agent execution pipeline. Table ~\ref{tab:root_cause_distribution} summarizes their frequency and distribution.

\begin{table}[htbp]
\centering
\caption{Distribution of Root Causes in Agentic Systems}
\label{tab:root_cause_distribution}
\begin{tabular}{lrr}
\toprule
\textbf{Root Cause Category} & \textbf{Count} & \textbf{Percentage} \\
\midrule
Data Schema Mismatches & 101 & 28.0\% \\
Dependency Drift & 79 & 21.9\% \\
State Management Complexity & 51 & 14.1\% \\
Model Interface Instability & 42 & 11.6\% \\
Error Propagation Failures & 29 & 8.0\% \\
Configuration Ambiguity & 15 & 4.2\% \\
External API Instability & 12 & 3.3\% \\
Architectural Coupling & 10 & 2.8\% \\
Resource Limit Violations & 7 & 1.9\% \\
Credential Misconfiguration & 5 & 1.4\% \\
Concurrency Issues & 5 & 1.4\% \\
Documentation Drift & 5 & 1.4\% \\
\bottomrule
\end{tabular}
\end{table}

\textbf{Data Schema Mismatches (101 occurrences):}  
This root cause involves the component boundaries where loosely structured producers — such as LLM outputs or tool return values — interact with strictly typed consumers, including parsers, validators, and downstream pipelines. It arises from implicit and often inconsistent assumptions about data representation across independently developed components. When these assumptions diverge, values are interpreted incorrectly or fail validation, which directly triggers \textit{Input Interpretation} faults. They may also propagate into \textit{Tool Invocation} faults when malformed data is passed to downstream services. In \textit{mindsdb} (\href{https://github.com/mindsdb/mindsdb/pull/4387}{PR \#4387}), missing values were incorrectly normalized during data processing, causing \texttt{NaN} and \texttt{pd.NA} errors and triggering a processing failure.

\textbf{Dependency Drift (79 occurrences):}  
This root cause involves the integration boundaries that connect agent frameworks with the external libraries, SDKs, and runtimes capable of independent evolution. These dependencies frequently introduce breaking changes, deprecate interfaces, or alter platform support, invalidating assumptions embedded within the framework. When these assumptions no longer hold, dependency resolution or execution environments become inconsistent, triggering \textit{Dependency and Environment Management} faults and, in some cases, \textit{Platform \& Integration Compatibility} faults. In \textit{crewAI} (\href{https://github.com/crewAIInc/crewAI/issues/1877}{Issue \#1877}), the framework's dependency on \texttt{uvloop} caused installation to fail on Windows because the library does not support that platform.

\textbf{State Management Complexity (51 occurrences):}  
This root cause involves the agent execution state, including context windows, tool traces, memory structures, and intermediate reasoning artefacts. Agent execution is inherently stateful and long-running, requiring multiple components to read from and write to a shared state. As the structure and semantics of this shared state evolve, there might be inconsistencies in how components interpret or update the state. This directly triggers \textit{Agent Lifecycle and State} faults, such as incorrect execution scheduling, inconsistent state transitions, or missing attributes during execution. In \textit{letta-ai/letta} (\href{https://github.com/letta-ai/letta/issues/1801}{Issue \#1801}), inconsistent handling of the \texttt{AgentState} structure caused runtime failures because different execution steps constructed \texttt{AgentState} with inconsistent fields, leading downstream components to access missing attributes and crash during state-dependent operations.

\textbf{Model Interface Instability (42 occurrences):}  
This root cause involves the boundary between agent frameworks and external model providers, where the framework translates reasoning requests into provider-specific API calls and interprets the resulting responses. It stems from assumptions about API parameters, response structures, and supported behaviours, which are implicitly encoded in request construction and parsing logic. When these assumptions are invalidated by API changes, they directly trigger \textit{LLM Integration Faults}, including incompatible API usage, rejected requests, incorrect parameter handling, or token accounting inconsistencies. In \textit{microsoft/autogen} (\href{https://github.com/microsoft/autogen/issues/1156}{Issue \#1156}), an update to the OpenAI client interface rendered the \texttt{request\_timeout} parameter unsupported, causing agent requests to fail.

\textbf{Error Propagation Failures (29 occurrences):}  
This root cause is present in the orchestration layers where failures originating from LLM calls, tool executions, or external service interactions might traverse multiple abstraction boundaries before becoming visible. Because errors or exceptions are not often handled appropriately across components, contextual information about the original failure is frequently lost, obscured, or replaced with generic messages. This loss of information directly triggers \textit{Failure Handling \& Implementation Robustness} faults and can also degrade \textit{Execution Monitoring} by obscuring the origin of failures. In \textit{Netflix/metaflow} (\href{https://github.com/Netflix/metaflow/pull/1444}{PR \#1444}), exceptions raised inside internal components were not propagated with sufficient context across execution boundaries.

\textbf{Configuration Ambiguity (15 occurrences):}  
This root cause involves the configuration assembled from multiple sources — such as environment variables, configuration files, constructor arguments, and defaults — without a clearly defined or consistently enforced schema. Different components or external services may interpret the same configuration parameters differently, particularly as schemas evolve or optional fields are handled inconsistently. These inconsistencies trigger \textit{External Access} faults, \textit{Tool Invocation} faults, and in some cases \textit{LLM Integration Faults}, as configuration values are interpreted differently across components. In \textit{pydantic/pydantic-ai} (\href{https://github.com/pydantic/pydantic-ai/issues/1798}{Issue \#1798}), a configuration mismatch between the model configuration and the MCP server caused agent execution to fail.

\textbf{External API Instability (12 occurrences):}  
This root cause involves the integration points where agent-generated actions are translated into calls against external APIs and services. External APIs evolve independently, and any changes to their endpoints, authentication mechanisms, or request and response schemas may not be reflected promptly in the framework’s integration logic. These changes directly trigger \textit{External Access} faults and may also lead to \textit{Tool Invocation} faults when request structures become incompatible. In \textit{langflow-ai/langflow} (\href{https://github.com/langflow-ai/langflow/pull/4386}{PR \#4386}), incorrect handling of the OpenAI client URL caused connection failures.

\textbf{Architectural Coupling (10 occurrences):}  
This root cause involves tight coupling between components and abstractions within the framework. As frameworks evolve incrementally, abstractions are introduced to simplify subsystem boundaries, which in turn introduce implicit dependencies among them. As these abstractions evolve, even small internal changes, such as modifying initialization logic, expected inputs, or execution flow, can break dependent components that rely on prior assumptions. This leads to inconsistent interactions across components, directly triggering \textit{Agent Lifecycle and State} faults (e.g., incorrect initialization or execution ordering) and \textit{System Coordination} faults (e.g., mismatched expectations between components). In \textit{crewAIInc/crewAI} (\href{https://github.com/crewAIInc/crewAI/issues/2263}{Issue \#2263}), introducing \texttt{CrewBase} as a decorator changed how agent classes were initialized, and components that relied on the previous construction pattern became incompatible, causing failures due to hidden dependencies on the abstraction’s internal behaviour.

\textbf{Resource Limit Violations (7 occurrences):}  
This root cause involves the violations of implicit assumptions about resource constraints such as memory usage, payload size, context length, or API rate limits. These limits are often not explicitly encoded or validated during development. When exceeded, they trigger \textit{Resource Interaction} faults and may also result in \textit{Platform \& Integration Compatibility} faults when execution environments cannot handle the input. In \textit{Arize-ai/phoenix} (\href{https://github.com/Arize-ai/phoenix/issues/7941}{Issue \#7941}), uploading CSV files containing large individual fields caused processing failures.

\textbf{Credential Misconfiguration (5 occurrences):}  
This root cause is present at the initialisation and integration points where the agent runtime establishes authenticated connections to external services. It arises because credentials are managed externally and are often not validated comprehensively at startup. Missing, invalid, or insufficiently scoped credentials therefore remain undetected until use, directly triggering \textit{External Access} faults during execution. In \textit{assafelovic/gpt-researcher} (\href{https://github.com/assafelovic/gpt-researcher/issues/909}{Issue \#909}), agent execution failed because the required API key was absent.

\textbf{Concurrency Issues (5 occurrences):}  
This root cause can be observed in orchestration components that manage asynchronous or parallel execution over shared mutable state. Without appropriate synchronisation mechanisms, concurrent operations introduce non-deterministic ordering of reads and writes. These conditions directly trigger \textit{System Coordination} faults and may also affect \textit{Agent Lifecycle and State} through inconsistent updates. In \textit{pydantic/pydantic-ai} (\href{https://github.com/pydantic/pydantic-ai/pull/2100}{PR \#2100}), a race condition during token refresh arose because the concurrent requests accessed shared authentication state without coordination.

\textbf{Documentation Drift (5 occurrences):}  
This root cause involves outdated documentation. Because documentation is maintained separately and often updated reactively, discrepancies emerge between documented usage and actual system behaviour. These discrepancies indirectly trigger \textit{Documentation Issues} faults and can propagate into other fault types when developers configure or use the system based on outdated assumptions. In \textit{crewAI} (\href{https://github.com/crewAIInc/crewAI/issues/213}{Issue \#213}), outdated documentation caused incorrect configuration and system behaviour.

\begin{tcolorbox}[colback=gray!10, colframe=black]
\textbf{Summary of RQ1:} Our analysis shows that faults in agentic AI systems arise from interactions among reasoning components, orchestration logic, external tools, data artefacts, and runtime environments. These faults surface as symptoms when intermediate data cannot be interpreted, workflows are interrupted, or errors propagate without proper handling. Their root causes are mainly inconsistent data formats, evolving dependencies, complex state management, and unstable model interfaces. Together, these findings show that agentic AI failures often originate at component boundaries where independently changing parts of the system must remain compatible.
\end{tcolorbox}

\subsection{RQ2: What statistically significant associations do exist among faults, symptoms, and root causes in agentic AI systems?}
\label{subsec:rq2}

\begin{figure}
    \centering
    \includegraphics[width=1\linewidth]{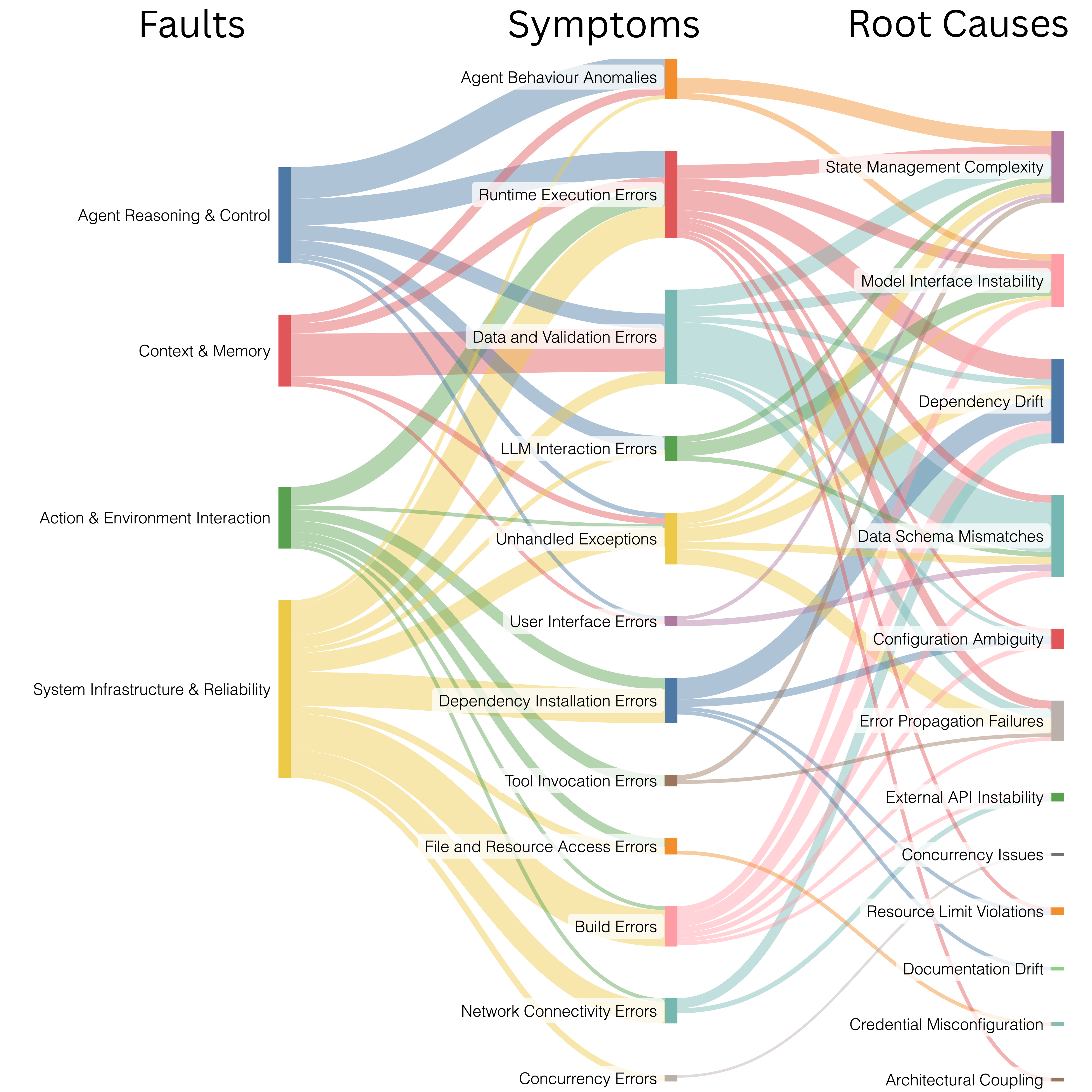}
    \caption{Failure propagation paths across architectural dimensions, observable symptoms, and root causes in agentic AI systems.}
    \label{fig:correlation}
\end{figure}

To understand how failures propagate in agentic AI systems, we identify high-confidence associations among fault types (\textit{Sub Category}), observed symptoms (\textit{Symptom Category}), and root cause categories (\textit{Root Cause Category}) using association rule mining. Each of the 385 labelled bug reports was encoded as a transaction capturing these elements, and rules were extracted using the Apriori algorithm~\cite{agarwal1994fast}. We evaluated the resulting rules using \textit{confidence}, retaining associations with confidence $\geq 0.30$, following existing literature~\cite{shah2025towards}. Our analysis shows that failures in agentic AI systems are not isolated events, but follow structured propagation patterns. Fig.~\ref{fig:correlation} visualizes these recurring pathways across architectural components, symptoms, and root causes.

\subsubsection{Relationship between the Fault Types and Symptoms}

\begin{table}[t]
\centering
\caption{Top five rules for Fault Type $\rightarrow$ Symptom}
\small
\begin{tabular}{llccc}
\toprule
\textbf{Fault Type} & \textbf{Symptom} & \textbf{Count} & \textbf{Confidence} \\
\midrule
UI Rendering Defect & User Interface Errors & 10 & 0.7692 \\
Exception Handling Failure & Unhandled Exceptions & 18 & 0.6667 \\
Structured Data Error & Data and Validation Errors & 28 & 0.6087 \\
Agent Step Execution Error & Runtime Execution Errors & 11 & 0.4231 \\
LLM Invocation Error & Runtime Execution Errors & 10 & 0.3571 \\
\bottomrule
\end{tabular}
\end{table}

Our derived rules provide important guidance for fault localization in agentic AI systems by linking observable symptoms to specific failure points in the execution pipeline. The rule \textit{UI Rendering Defect $\rightarrow$ User Interface Errors} exhibits high confidence (0.7692), meaning that 76.92\% of UI rendering defects manifest as user interface errors. This indicates that when UI errors are observed, they are frequently attributable to rendering defects rather than upstream components. In agentic systems, UI components act as terminal layers for presenting multi-step outputs, tool interactions, and intermediate reasoning traces generated by the agent. As a result, these errors typically originate at the presentation layer and reflect failures in rendering structured outputs, rather than issues in reasoning, planning, or tool execution. Developers can therefore prioritize inspection of rendering components before tracing upstream execution. The rule \textit{Exception Handling Failure $\rightarrow$ Unhandled Exceptions} (confidence 0.6667) reflects failures in managing execution errors across agent steps. In agentic systems, execution spans planning, tool invocation, and intermediate reasoning stages. When exception handling is poor, failures originating in earlier steps could propagate through the orchestration loop and surface as unhandled exceptions at later steps. This suggests that developers should focus on enforcing appropriate error-handling practices to prevent error propagation across agent components. 

The rule \textit{Structured Data Error $\rightarrow$ Data and Validation Errors} has strong confidence (0.6087), indicating that data-related failures are consistently observable at validation stages. In agentic systems, LLM outputs are often transformed into structured inputs to downstream tools. Thus, the mismatches between generated outputs and expected schemas occur frequently. Appropriate validation should be applied immediately after data generation steps to prevent error propagation. Unlike the above single fault-to-symptom mappings, \textit{Runtime Execution Errors} are associated with multiple fault types, including \textit{Agent Step Execution Error} (confidence 0.4231) and \textit{LLM Invocation Error} (confidence 0.3571). Their lower confidence values indicate that runtime errors are not uniquely tied to a single fault type. Instead, runtime execution acts as an aggregation point where faults originating from different components become observable only after propagating through multiple steps. This reflects tightly coupled orchestration in the agentic systems, where failures are delayed and require step-wise tracing across the execution pipeline.

\subsubsection{Relationship between the Symptoms and Root Causes}

\begin{table}[t]
\centering
\caption{Top five rules for Symptom $\rightarrow$ Root Cause}
\small
\begin{tabularx}{\linewidth}{l l c c}
\toprule
\textbf{Symptom} & \textbf{Root Cause} & \textbf{Count} & \textbf{Confidence} \\
\midrule
Dependency Installation Errors & Dependency Drift & 26 & 0.9286 \\
Build Errors & Dependency Drift & 18 & 0.5625 \\
Agent Behaviour Anomalies & State Management Complexity & 12 & 0.5000 \\
LLM Interaction Errors & Model Interface Instability & 10 & 0.4545 \\
Unhandled Exceptions & Error Propagation Failures & 19 & 0.4222 \\
\bottomrule
\end{tabularx}
\end{table}

Our symptom-to-root-cause rules provide guidance for identifying underlying root causes of a failure once a symptom is observed. The rule \textit{Dependency Installation Errors $\rightarrow$ Dependency Drift} has very high confidence (0.9286), indicating that installation failures are often caused by inconsistencies in dependency versions or environment configuration. These failures prevent environment setup and successful installation, underscoring the need for clear dependency specifications, version pinning, and reproducible environments. Similarly, the rule \textit{Build Errors $\rightarrow$ Dependency Drift} (confidence 0.5625) shows that dependency inconsistencies also affect the build stage, leading to build failures. The rule \textit{Agent Behaviour Anomalies $\rightarrow$ State Management Complexity} (confidence 0.5000) indicates that issues in state handling are a primary contributor to anomalous agent behaviours. In agentic systems, state evolves across the steps, and failures in updating, saving, or retrieving state can lead to anomalies in execution in the case of anomalous agent behaviours. This suggests that diagnosis should focus on state representation and transitions across components. The rule \textit{LLM Interaction Errors $\rightarrow$ Model Interface Instability} (confidence 0.4545) captures failures at the interface between the agent and the language model. These errors arise when expected input or output formats are not consistently enforced, indicating defects in prompt construction, response parsing, and interface validation mechanisms. Finally, the rule \textit{Unhandled Exceptions $\rightarrow$ Error Propagation Failures} (confidence 0.4222) shows that unhandled exceptions are often the result of failures in managing errors across execution steps. In multi-step pipelines, errors originating in earlier components could propagate without validation and surface as exceptions, suggesting that structured error handling and validation should be enforced at each step.

\subsubsection{Key Insights}

The above rules show that failures in agentic AI systems are often shaped by agents' orchestration, autonomy, and environment dependence. High-confidence rules explain how error symptoms correspond to the types of faults or root causes. In contrast, shared symptoms such as runtime execution errors indicate fault propagation across components. Such a presence of many-to-one mappings indicates delayed failure manifestation, where faults introduced in earlier steps surface only at later stages due to multi-step orchestration. On the other hand, autonomy introduces implicit assumptions about data formats, interface contracts, and state consistency. Violations of these assumptions lead to recurring patterns observed in structured data errors, LLM interaction errors, and state-related anomalies. These findings suggest that robustness requires early validation to limit orchestration-related fault propagation, explicit interface contracts and state management to handle autonomy-related assumptions, and reproducible environments to control environment-dependent failures.

\begin{tcolorbox}[colback=gray!10, colframe=black]
\textbf{Summary of RQ2:} Failures in agentic AI systems follow structured propagation patterns shaped by orchestration, autonomy, and environment dependence. Multi-step orchestration causes early-introduced faults to propagate and surface later as shared runtime errors or exceptions. Autonomy introduces implicit assumptions about data formats, interface contracts, and state, leading to recurring failures related to structured data, interactions, and state. Environment dependence contributes to failures tied to dependencies and configuration. As a result, failures that propagate across steps are harder to trace back to their origin, whereas failures tied to specific components tend to produce clearer and more localized symptoms.
\end{tcolorbox}

\definecolor{col1}{RGB}{88,28,92}
\definecolor{col2}{RGB}{31,50,120}
\definecolor{col3}{RGB}{30,130,140}
\definecolor{col4}{RGB}{60,170,80}
\definecolor{col5}{RGB}{235,205,40}

\subsection{RQ3: To what extent do the derived taxonomies of bugs, symptoms, and root causes align with the practical experiences of agentic system developers?}
\label{sec:rq3}

To evaluate the ecological validity of our proposed taxonomy, we conducted a structured survey of software developers actively building or maintaining agentic AI systems. Participants evaluated the practical relevance of each third-level fault type and identified potential omissions or mis-specifications. We present a mix of quantitative measures and qualitative feedback reflecting the real-world relevance of our findings as follows.

\subsubsection{Quantitative Validation Results}
\label{subsec:rq3_quant}

From 145 participants, we collected relevance ratings for the 34 third-level fault types, which correspond to the leaf nodes of the taxonomy. Ratings were collected using a five-point Likert scale (1 = not relevant, 5 = highly relevant). The overall mean rating was 3.97 (median = 4.0, SD = 0.87), indicating that participants generally perceived the fault types as relevant in practice. Overall, 74.9\% of responses rated fault types as 4 or higher, while 4.4\% rated them as 2 or lower. A one-sample Wilcoxon signed-rank test against the neutral midpoint of 3.0 showed that participants' ratings were significantly higher than neutral ($M = 3.97$, $Mdn = 4.03$, $W = 10381.5$, $p < 0.001$). Internal consistency across fault-type ratings was high (Cronbach's $\alpha = 0.91$), indicating that participants evaluated the fault types consistently. We analyze the participant ratings at two levels of the taxonomy hierarchy. For dimension-level results, we aggregate the 34 fault-type ratings under their corresponding first-level dimensions.

\textbf{Dimension-Level Endorsement.}
All four first-level dimensions received consistent ratings from participants, with mean ratings ranging from 3.93 to 4.04. In the survey, participants rated the practical relevance of the third-level fault types associated with each first-level dimension. \emph{Tooling, Integration \& Actuation} received the highest mean rating (4.04), followed by \emph{Agent Reasoning \& Control} (4.00), \emph{Context \& Memory} (3.94), and \emph{System Infrastructure \& Reliability} (3.93). These results indicate that participants rated faults across all four major aspects of agentic AI systems as relevant in practice.  Figure~\ref{fig:rq3_dimensions} shows the distribution of ratings for the four dimensions (D1: Agent Reasoning \& Control; D2: Context \& Memory; D3: Tooling, Integration \& Actuation; D4: System Infrastructure \& Reliability).

\begin{figure}[h]
\centering
\begin{tikzpicture}
\begin{axis}[
    xbar stacked,
    scale only axis,
    width=0.72\linewidth,
    height=3.0cm,
    xmin=0, xmax=100,
    bar width=14pt,
    xlabel={Percentage (\%)},
    symbolic y coords={D1,D2,D3,D4},
    ytick=data,
    y dir=reverse,
    enlarge y limits=0.22,
    axis x line*=bottom,
    axis y line*=left,
    xmajorgrids=false,
    xtick={0,20,40,60,80,100},
    tick label style={font=\scriptsize},
    label style={font=\scriptsize},
    legend style={
        at={(0.5,-0.32)},
        anchor=north,
        legend columns=5,
        draw=none,
        font=\scriptsize,
        /tikz/every even column/.append style={column sep=0.35cm}
    },
    clip=false
]
\addplot+[xbar,draw=none,fill=col1] coordinates {(2.0,D1)(2.4,D2)(1.8,D3)(1.6,D4)};
\addplot+[xbar,draw=none,fill=col2] coordinates {(2.1,D1)(2.4,D2)(2.6,D3)(2.8,D4)};
\addplot+[xbar,draw=none,fill=col3] coordinates {(21.0,D1)(21.7,D2)(17.0,D3)(22.9,D4)};
\addplot+[xbar,draw=none,fill=col4] coordinates {(44.1,D1)(46.2,D2)(47.3,D3)(46.7,D4)};
\addplot+[xbar,draw=none,fill=col5] coordinates {(30.8,D1)(27.3,D2)(31.3,D3)(26.0,D4)};
\vspace{0.2em}
\legend{1 -- Very Low, 2 -- Low, 3 -- Moderate, 4 -- High, 5 -- Very High}
\end{axis}
\end{tikzpicture}
\caption{Distribution of practitioner relevance across the four dimensions.}
\label{fig:rq3_dimensions}
\end{figure}

\textbf{Fault-Type-Level Agreement.}
We next analyze agreement for the individual third-level fault types, which are the leaf nodes of the taxonomy and were used as survey items. We summarize fault-type-level agreement using the proportion of participants who assigned a relevance rating of 4 or 5 on the 5-point scale. Overall, 33 of the 34 fault types (97.1\%) received a rating $>=$ 4 from a majority of participants, indicating that most leaf-level fault types were viewed as relevant by practitioners. Moreover, 25 fault types (73.5\%) received ratings of $>=$ 4 from at least 70\% of participants, and 31 fault types (91.2\%) received ratings of $>=$ 4 from at least 60\% of participants. The highest-scoring fault types were related to LLM invocation, tool configuration, tool invocation, execution monitoring, package resolution, exception handling, and implementation logic. These results suggest that practitioners rated faults at system boundaries and execution interfaces as especially relevant, particularly faults involving interactions among LLM reasoning modules, external tools, and runtime environments. The only fault type for which fewer than half of the participants assigned a rating of 4 or 5 was \emph{UI Rendering Defect} (46.2\%; mean = 3.54). This lower score suggests that practitioners perceived UI-related failures as less central to agentic system reliability than failures involving LLM invocation, tool use, execution monitoring, and runtime dependencies. This likely reflects the deployment context of many agentic systems, which are exposed through APIs, scripts, or automated workflows rather than interactive user interfaces. Thus, UI-related failures remain relevant for interface-driven agents, but they may be less frequently encountered by practitioners working on backend or tool-mediated agent pipelines.

\textbf{Variation by Developer Experience.}
According to our analysis, the perceived relevance of the proposed fault taxonomy was consistent across experience levels. As shown in Fig.~\ref{fig:rq3_experience}, participants with more than one year of experience reported a mean rating of 3.91, with 72.6\% of ratings at 4 or 5. Participants with six months to one year of experience reported a mean rating of 4.00, with 75.3\% of ratings at 4 or 5. Participants with less than six months of experience reported the highest mean rating of 4.13, with 81.1\% of ratings at 4 or 5. These results indicate that our taxonomy captures failure modes observed across different levels of expertise.

\begin{figure}[h]
\centering
\begin{tikzpicture}
\begin{axis}[
    xbar,
    scale only axis,
    width=0.68\linewidth,
    height=2.2cm,
    xmin=3.6, xmax=4.3,
    bar width=13pt,
    xlabel={Mean Relevance Rating},
    symbolic y coords={$>$12 months, 6 mo.--1 yr., $<$6 months},
    ytick=data,
    y dir=reverse,
    enlarge y limits=0.38,
    axis x line*=bottom,
    axis y line*=left,
    xmajorgrids=true,
    xtick={3.6,3.7,3.8,3.9,4.0,4.1,4.2,4.3},
    tick label style={font=\footnotesize},
    label style={font=\footnotesize},
    nodes near coords,
    nodes near coords align={horizontal},
    every node near coord/.append style={
        font=\footnotesize\bfseries,
        text=col3,
        xshift=6pt,
        /pgf/number format/fixed,
        /pgf/number format/precision=2
    },
    clip=false
]
\addplot[xbar, draw=none, fill=col3!80] coordinates {
    (3.91,$>$12 months)
    (4.00,6 mo.--1 yr.)
    (4.13,$<$6 months)
};
\end{axis}
\end{tikzpicture}
\caption{Mean relevance ratings by developer experience level with agentic AI systems. Differences across groups are non-significant ($p > 0.05$), indicating consistent relevance.}
\label{fig:rq3_experience}
\end{figure}

\subsubsection{Qualitative Validation of Taxonomy}
\label{subsec:rq3_qual}

In addition to numerical ratings, participants provided open-ended feedback on missing, unclear, or difficult-to-apply failure types. Most participants (122/145; 83.8\%) indicated that the proposed taxonomy broadly reflects their experience with agentic systems. In particular, they appreciated the layered structure of the taxonomy, especially its separation of failures involving LLM reasoning, tool integration, and infrastructure components. The following box reports representative positive feedback.

\begin{tcolorbox}[colback=green!6,colframe=green!50!black,title={Representative Positive Practitioner Feedback}]
\small 

\emph{“The taxonomy captures most of the failure types we encounter when building agent systems. It clearly reflects practical development challenges.”} (R12)

\emph{“I like how the categories separate reasoning issues from infrastructure and tool integration problems.”} (R18)

\emph{“This structure reflects the real debugging process in agent pipelines, where issues can arise from model behaviour, tools, or infrastructure.”} (R27)

\emph{“The taxonomy is helpful because it highlights where failures occur in the system rather than just describing symptoms.”} (R34)

\emph{“Many of the categories map directly to problems we see when integrating LLMs with external APIs and tools.”} (R45)

\emph{“The separation between cognition, perception, tooling, and environment makes the failure landscape easier to reason about.”} (R63)

\emph{“This provides a helpful mental model for analysing failures in complex agent workflows.”} (R70)

\emph{“The taxonomy aligns well with the types of debugging problems we face in production systems.”} (R96)

\end{tcolorbox}

In the remaining section, we report recurring themes from the open-ended survey responses. Participants were asked to provide additional feedback on the taxonomy, including categories that were unclear, missing, or difficult to apply. We analyzed these responses using grounded theory coding and clustered related comments into recurring themes.

\textbf{Semantic Failures in LLM Reasoning.}
In addition to our four agent dimensions in the proposed taxonomy, one survey respondent (R23) suggested differentiating between syntactic and semantic failures. Syntactic failures violate expected formats, schemas, or interface contracts and often fail during parsing, validation, or execution, whereas semantic failures occur when outputs are structurally valid but logically incorrect for the intended task. For example, agents may report successful task completion even when compilation or validation checks for the output fail. Such failures can propagate silently through the reasoning chains and often require explicit verification mechanisms. This theme is partly related to \emph{LLM Integration Faults} and \emph{Input Interpretation}, but the response suggests that semantic correctness may require a clearer distinction from format, schema, or parsing failures.

\begin{commentbox}
“Many failures we see are not syntax errors but cases where the LLM output looks correct but is logically wrong.” (R23)
\end{commentbox}

\textbf{Multi-Agent Coordination Failures.}
Several participants (e.g., R41) reported failure modes specific to multi-agent environments, including misinterpretation of inter-agent messages, cascading prompt injection across agents, circular task dependencies, and coordination breakdowns. In these cases, the failure emerges from how autonomous agents interpret, transform, and act on shared task context across multiple steps, rather than from a fixed sequence of object or method calls. A message may be syntactically valid but still alter another agent's goal, trigger an unsafe delegation, or propagate incorrect assumptions through the workflow. These observations suggest the need to distinguish low-level synchronisation problems, such as race conditions or shared-state conflicts, from agent-level coordination failures involving task delegation, message interpretation, and propagation of invalid assumptions. This theme is partly related to \emph{Agent Lifecycle and State} and \emph{System Coordination}, but the responses suggest that coordination among autonomous agents may need more explicit treatment in future refinements.

\begin{commentbox}
“In multi-agent systems the biggest problems come from coordination between agents rather than individual tool failures.” (R41)
\end{commentbox}

\textbf{Human-in-the-Loop Workflow Failures.}
Systems incorporating approval gates or verification checkpoints may fail when workflow control logic is implemented incorrectly. Reported examples (R52) include stalled workflows awaiting human input, due to incorrect escalation logic, and failures to enforce verification policies where needed. This theme overlaps with \emph{Agent Lifecycle and State}, \emph{Execution Monitoring}, and \emph{Failure Handling \& Implementation Robustness}, but the responses suggest that human approval and verification checkpoints may need to be represented more explicitly.

\begin{commentbox}
“We sometimes see workflows get stuck waiting for human approval because the orchestration logic fails to trigger the next step.” (R52)
\end{commentbox}

\textbf{Observability Limitations in Agent Pipelines.}
Participants also emphasised the need for improved observability mechanisms to debug agent pipelines. They highlighted the importance of structured execution traces capturing reasoning steps, tool invocations, and state transitions across agents. This theme is directly related to \emph{Execution Monitoring}, which covers defects in logging, tracing, and metric instrumentation. However, the responses suggest that observability may also be a cross-cutting diagnostic concern across reasoning, tool use, and state transitions.

\begin{commentbox}
“Debugging agent pipelines is difficult because we cannot easily see the intermediate reasoning steps or tool interactions.” (R57)
\end{commentbox}

\textbf{Behavioural Resource Exhaustion.}
Several participants described failures caused by inefficient orchestration behaviour, such as agents continuing execution after achieving goals, excessive polling loops, or runaway API usage. These observations highlight the importance of resource-aware orchestration mechanisms. This theme overlaps with \emph{Agent Lifecycle and State} and LLM context-window or token-accounting faults under \emph{LLM Integration Faults}, but the responses suggest a clearer distinction for resource exhaustion caused by agent behaviour, such as repeated actions, unnecessary tool calls, or uncontrolled execution loops.

\begin{commentbox}
“Agents sometimes keep running even after solving the task, repeatedly calling APIs until they hit rate limits.” (R72)
\end{commentbox}

\begin{tcolorbox}[colback=gray!10,colframe=black]
\textbf{Summary of RQ3:} A practitioner survey involving 145 software developers supports the practical relevance of the proposed taxonomy. Across 4,930 ratings, the taxonomy achieved a mean relevance score of 3.97 out of 5, with 74.9\% of responses assigning a rating of 4 or higher. Qualitative feedback further supports the taxonomy's coverage while identifying opportunities for refinement related to semantic reasoning failures, multi-agent coordination, human approval workflows, observability, and resource-aware orchestration.
\end{tcolorbox}

\section{Discussion}
\label{sec:discuss}

Our analysis of 385 faults from 40 agentic AI repositories reveals that failures in agentic systems arise from the interaction between probabilistic LLM-based reasoning and deterministic software components. These failures rarely remain localized; instead, they propagate across reasoning pipelines, tool invocation mechanisms, and runtime environments. In this section, we discuss the implications of our findings for key stakeholders involved in the design, development, and deployment of agentic systems.

\subsection{Implications for System Designers}

The hybrid nature of agentic systems introduces failures across interactions among the language-model core, tool APIs, external services, runtime environments, and structured data artifacts. Designers should account for these failures when architecting agent pipelines. In particular, interfaces between LLM outputs and downstream components should be treated as first-class design components, with clearly defined schemas, validation layers, and fallback mechanisms. The prevalence of data and validation errors (Table 1) and data schema mismatches (Table 2) suggests that loosely specified interfaces are a primary source of system fragility. Designing agents with strongly typed intermediate representations and explicit contracts can reduce ambiguity and prevent failures from propagating across the pipeline. Additionally, modular architectures that isolate agents' reasoning, state management, and execution can limit the spread of faults and improve system robustness.

\subsection{Implications for Developers}

Our findings encourage the developers to rethink debugging and validation practices. Traditional debugging approaches predominantly assume deterministic execution, but agentic systems exhibit non-deterministic behaviour due to LLM outputs. As a result, failures may not be consistently reproducible and may manifest only under specific execution contexts. To enhance debugging, developers should incorporate validation checks at multiple stages of the agent pipeline, particularly after LLM outputs and before tool invocation. The observed propagation patterns (Fig. 3) indicate that errors often originate early but surface later as runtime exceptions or anomalous behaviour. Early validation of intermediate outputs can therefore prevent cascading failures. Furthermore, given that different faults of an agentic system can show similar external symptoms (Table 3), developers should avoid relying solely on observed behaviour for diagnosis. Instead, debugging workflows should incorporate structured tracing of reasoning steps, state transitions, and tool interactions to accurately identify root causes.

\subsection{Implications for Infrastructure and Tooling Providers}

The high prevalence of dependency drift and integration-related failures (Table 2) underscores the fragility of the agent ecosystem. Tooling and infrastructure providers should prioritize stability and backward compatibility in APIs, as even minor changes can disrupt downstream agent pipelines. Existing AgentOps and LLM observability platforms (e.g., AgentOps~\cite{dong2024agentops}, LangSmith~\cite{langsmithobservability2026}) already provide support for tracing prompts, model responses, tool invocations, costs, latency, and multi-agent interactions.

However, our findings suggest the need for more targeted debugging support for agentic systems, especially multi-agent workflows. Infrastructure support should move beyond recording execution events to correlate traces across agents, link tool failures to upstream reasoning or state transitions, and support the replay of failure-inducing execution paths. Such capabilities would help developers distinguish local tool or API failures from failures that propagate through task delegation, shared context, or inter-agent communication. Testing support should also account for multi-agent execution conditions. Existing frameworks such as Maia~\cite{maiaframework2025} provide pytest-based support for testing multi-agent AI systems, but our taxonomy suggests additional test dimensions: simulation of external dependency failures, controlled variability in LLM outputs, and checks for circular delegation or non-terminating workflows. These capabilities are needed to identify failures that emerge only in multi-step, tool-mediated, or multi-agent executions.

\subsection{Implications for Reliability Engineering}

The identified propagation patterns suggest that failures in agentic systems should be treated as multi-stage processes rather than isolated defects (Sections 3.1 and 3.2). Reliability engineering practices should therefore focus on detecting and mitigating failures early in the pipeline before they propagate. For example, automated monitors can be designed to detect common failure signatures, such as outputs that violate expected schemas, state updates that conflict with predefined state invariants, or repeated tool invocation failures. Similarly, mechanisms for state verification and rollback can help prevent faulty states from affecting subsequent reasoning steps. The distinction between internal and boundary-related failures also has implications for mitigation strategies. Internal failures, such as those related to state management or error propagation, require improved validation and tracing, while boundary failures, such as API instability or resource constraints, can be mitigated through redundancy, retries, and graceful degradation strategies.

\subsection{Implications for the Agentic AI Research Community}

From a research perspective, our findings suggest that agentic AI failures should be studied as pipeline-level phenomena rather than isolated LLM errors. Our taxonomy provides a structured basis for this analysis by identifying where failures occur across reasoning, tool invocation, state management, integration, and execution environments. Researchers can use these categories to annotate failure datasets, compare agent frameworks, and report benchmark results using consistent labels. The association rules further show how faults, symptoms, and root causes co-occur, making the taxonomy useful for studying failure propagation rather than only failure frequency. These rules can guide testable hypotheses about where failures originate, how they surface, and which intermediate artifacts should be inspected during evaluation. The developer survey provides an external check on the taxonomy's practical relevance. Practitioners rated the categories as relevant to development and debugging work, which increases confidence that the taxonomy reflects failures encountered in practice. This does not establish completeness, but it supports the taxonomy as a starting point for empirical studies, benchmark annotation, and reliability-oriented tooling. Overall, future research should evaluate agentic systems using traceable and propagation-aware methods that capture multi-step reasoning, tool calls, state updates, and environment dependencies. Agentic AI systems should therefore be studied as hybrid systems whose reliability depends on both probabilistic reasoning and deterministic infrastructure.

\section{Threats to Validity}
\label{sec:threat}

\textbf{Threats to \emph{internal validity}} arise from potential biases in manual analysis and automated filtering. Manual annotation of fault types, symptoms, and root causes may introduce subjectivity. We mitigated this risk through batch-wise coding, iterative refinement of the taxonomy, and regular consensus discussions to resolve disagreements. Another threat concerns the automated filtering of issues using GPT-4.1. To evaluate its reliability, we compared model predictions against a human-labelled ground truth, obtaining 83\% accuracy and 97\% recall. We further manually inspected a representative subset of predictions to confirm consistency. Finally, association rule mining in RQ2 may produce spurious correlations. To reduce this risk, we retained only rules with confidence $\geq 30\%$, following existing literature~\cite{shah2025towards}, and interpreted the resulting rules as high-confidence co-occurrence patterns rather than causal relationships.

\textbf{Threats to \emph{construct validity}} concern whether the dataset and taxonomy accurately capture real-world faults in agentic AI systems. GitHub issue reports may contain incomplete descriptions or speculative diagnoses. To reduce noise, we retained only issues with evidence of resolution, such as a closed status, a linked merged pull request, or a discussion confirming that the fault had been fixed. Another potential threat stems from the completeness of the taxonomy. We mitigated this through an iterative coding process using 385 faults from 40 agent systems and a practitioner survey of developers with varying experience levels. The survey results indicated strong alignment between the taxonomy and practitioner experience while identifying several areas for refinement.

\textbf{Threats to \emph{external validity}} relate to the generalisability of our findings. The study focuses on Python-based repositories, which may limit applicability to other programming ecosystems. However, Python currently dominates agentic AI development due to the availability of major LLM frameworks and tooling. Our dataset is also limited to GitHub repositories, which may not fully represent industrial deployments or proprietary systems. To improve representativeness, we employed stratified sampling across four repository types (frameworks, libraries, tools, and applications), as per existing literature~\cite{kim2021denchmark} and restricted selection to repositories with more than 1{,}000 stars and at least 30 issues, favouring mature and actively maintained projects. To ensure transparency and reproducibility, we have made our dataset and replication package publicly available~\cite{replicationpackage}.

\section{Related Work}
\label{sec:related}

\textbf{Taxonomies of failures in agentic AI systems.}
The growing adoption of agentic workflows has motivated several efforts to categorise failures in such systems. Existing taxonomies typically adopt one of three perspectives: platform-centric, interaction-centric, or task-centric. From a platform-centric viewpoint, Ma et al.~\cite{ma2025diagnosing} classify failures originating from orchestration infrastructures that manage agent lifecycles, showing how platform-level constraints can propagate into system-wide crashes. This perspective explains failures tied to agent execution platforms but provides limited coverage of faults originating in reasoning, state management, tool use, data handling, or external dependencies. In multi-agent environments, Cemri et al.~\cite{cemri2025multi} adopt an interaction-centric perspective, analysing emergent failures such as coordination loops and hallucinated communication protocols that arise during inter-agent communication. Their taxonomy captures failures specific to agent-agent interaction, but it does not systematically connect these failures to implementation-level root causes or to broader system components. Task-centric approaches, including those proposed by Lu et al.~\cite{lu2025} and Rahardja et al.~\cite{rahardja2025can}, categorise failures based on task outcomes or execution traces. In particular, Rahardja et al.~\cite{rahardja2025can} focus on logic errors and syntax violations encountered by self-debugging agents. These approaches are useful for evaluating task success and repair behaviour, but they primarily describe what went wrong during task execution rather than where faults originate within the agent architecture and how they propagate across components. More recently, Ning et al.~\cite{ning2026defining} examine defects in LLM-based autonomous agents from a code analysis perspective. By mining developer discussions and issues on Stack Overflow and GitHub, they define eight categories of agent code defects, including errors in tool invocation, LLM output parsing, and API interactions, and propose a static analysis framework to automatically detect these defects in source code. While their work provides valuable insights into implementation-level defects within single-agent workflows, its focus is primarily on detecting specific code-level issues rather than analysing how faults emerge, manifest, and propagate across the broader architecture of agentic systems. In contrast, our work adopts a component-grounded perspective that links faults to agentic system components and jointly analyses fault types, observable symptoms, and root causes. This perspective integrates platform, interaction, task, and implementation concerns into a unified taxonomy and complements it with association rules that reveal recurring propagation patterns among faults, symptoms, and causes.

\textbf{Failure attribution and diagnosis in agentic systems.}
Recent research has also investigated mechanisms for diagnosing failures in agent workflows. Zhang et al.~\cite{zhang2025agent} propose a voting-based consensus mechanism to identify which agent is responsible for a failure in multi-agent tasks, addressing the challenge of blame attribution in non-deterministic environments. Zhu et al.~\cite{zhu2025llm} explore self-reflective feedback loops in which agents analyse execution traces to refine prompts and reasoning strategies. These approaches aim to improve system robustness and debugging workflows by enabling agents to detect or correct failures during execution. However, they largely treat agents as black-box components and focus on behavioural attribution or prompt refinement rather than analysing how failures arise from interactions among agent components, tool calls, shared state, external services, and non-deterministic LLM outputs.

\textbf{Positioning of this work.}
This work complements existing research through a \emph{component-grounded failure characterization and propagation analysis} of agentic AI systems. Instead of analysing failures solely in terms of task outcomes or agent behaviour, we adopt a component-grounded perspective that captures agent functionality, memory subsystems, planning and control logic, tool interfaces, and runtime environments. Our study analyses 385 real-world faults collected from 40 frameworks, libraries, tools, and applications, providing coverage across multiple layers of the agentic AI software stack. By mining high-confidence associations between root causes, fault types, and symptoms, we uncover recurring fault propagation patterns within agent pipelines. Finally, unlike earlier studies~\cite{cemri2025multi, rahardja2025can, ning2026defining}, we validate the resulting taxonomy through a large-scale practitioner survey, which reports strong support and useful suggestions for further refinement.

\section{Conclusion and Future Work}
\label{sec:conclusion}

Agentic AI systems exhibit distinctive failure modes arising from the interaction of probabilistic LLM behaviour, autonomous control loops, tool-mediated actuation, and rapidly evolving software ecosystems. This paper presented a large-scale, architecture-grounded empirical study of faults in agentic AI systems. We collected 13{,}602 closed issues and merged pull requests from 40 widely used repositories and conducted in-depth manual analysis of 385 representative faults. Using a grounded-theory approach, we derived taxonomies of fault types, symptoms, and root causes and mapped them to key architectural components of agentic systems, including cognition and orchestration, tool actuation, context and memory management, execution environments, and system reliability.

Our findings show that failures in agentic systems rarely remain confined to a single component. Instead, they frequently emerge at the boundaries between LLM reasoning and deterministic program logic, propagate through stateful control loops, and are amplified by fragile integration with external tools and evolving software dependencies. These patterns reveal structural differences between faults in agentic systems and those observed in traditional software. Our findings highlight several implications for the engineering of reliable agentic systems. First, robust interfaces are needed at the boundary between model outputs and programmatic components to support safe and predictable tool invocation. Second, long-running agents require systematic mechanisms to manage and validate their internal state across reasoning steps. Third, agent pipelines should reduce exposure to dependency and external service failures by pinning package versions, checking API contracts at runtime, isolating tool calls behind adapters, and defining fallback behaviours for outages, rate limits, authentication failures, and response format changes.

Future work will extend this study along several directions. We plan to broaden the dataset to include additional programming languages, deployment environments, and industrial systems. We will investigate stronger model-code interface contracts and validation mechanisms that reduce errors when translating LLM outputs into structured program actions. We also aim to develop improved observability and diagnostic techniques for agent pipelines, including structured execution traces and automated debugging support. By systematically characterising fault types, symptoms, and propagation pathways in real-world agentic AI software, this work provides empirical foundations for improving the reliability, diagnosability, and engineering practices of agentic AI systems.

\bibliographystyle{ACM-Reference-Format}
\bibliography{main}
\end{document}